\documentclass[preprint2]{proto}
\usepackage{natbib}
\bibpunct{(}{)}{;}{a}{,}{,}
\usepackage{times}
\usepackage{color}
\usepackage{amsmath}

\begin{document}

\title{\textbf{\LARGE Planet-Disc Interactions and Early Evolution of Planetary Systems}}

\author {\textbf{\large Cl{\'e}ment Baruteau}}
\affil{\small\em University of Cambridge}  
\author {\textbf{\large Aur{\'e}lien Crida}}
\affil{\small\em Universtit\'e Nice Sophia Antipolis / Observatoire de la C{\^o}te d'Azur}
\author {\textbf{\large Sijme-Jan Paardekooper}}
\affil{\small\em University of Cambridge}
\author {\textbf{\large Fr{\'e}d{\'e}ric Masset}}
\affil{\small\em Universidad Nacional Aut{\'o}noma de M{\'e}xico}
\author {\textbf{\large J{\'e}r{\^o}me Guilet}}
\affil{\small\em University of Cambridge}
\author {\textbf{\large Bertram Bitsch}}
\affil{\small\em Universtit\'e Nice Sophia Antipolis / Observatoire de la C{\^o}te d'Azur}
\author {\textbf{\large Richard Nelson}}
\affil{\small\em Queen Mary, University of London}
\author {\textbf{\large Wilhelm Kley}}
\affil{\small\em Universit{\"a}t T{\"u}bingen}
\author {\textbf{\large John Papaloizou}}
\affil{\small\em University of Cambridge}

\begin{abstract}
  \baselineskip = 11pt \leftskip = 0.65in \rightskip =
  0.65in \parindent=1pc {\small The great diversity of extrasolar
    planetary systems has challenged our understanding of how planets
    form, and how their orbits evolve as they form. Among the various
    processes that may account for this diversity, the gravitational
    interaction between planets and their parent protoplanetary disc
    plays a prominent role in shaping young planetary systems.
    Planet-disc forces are large, and the characteristic times for the
    evolution of planets orbital elements are much shorter than the
    lifetime of protoplanetary discs. The determination of such forces
    is challenging, because it involves many physical mechanisms and
    it requires a detailed knowledge of the disc structure.  Yet, the
    intense research of the past few years, with the exploration of
    many new avenues, represents a very significant improvement on the
    state of the discipline. This chapter reviews current
    understanding of planet-disc interactions, and highlights their
    role in setting the properties and architecture of observed
    planetary systems.  \\~\\~\\~}
 \end{abstract}

\section{\textbf{INTRODUCTION}}
\noindent 
Since the fifth edition of Protostars and Planets in 2005 (PPV) the
number of extrasolar planets has increased from about 200 to nearly
1000, with several thousand transiting planet candidates awaiting
confirmation. These prolific discoveries have emphasized the amazing
diversity of exoplanetary systems. They have brought crucial
constraints on models of planet formation and evolution, which need to
account for the many flavors in which exoplanets come. Some giant
planets, widely known as the hot Jupiters, orbit their star in just a
couple of days, like 51 Pegasus b \citep{MQ95}. Some others orbit
their star in few ten to hundred years, like the four planets known to
date in the HR~8799 planetary system \citep{Marois10}. At the time of
PPV, it was already established that exoplanets have a broad
distribution of eccentricity, with a median value $\approx 0.3$
\citep{Takeda05}. Since then, measurements of the Rossiter McLaughlin
effect in about 50 planetary systems have revealed several hot
Jupiters on orbits highly misaligned with the equatorial plane of
their star \citep[e.g.,][]{Albrecht12}, suggesting that exoplanets
could also have a broad distribution of inclination. Not only should
models of planet formation and evolution explain the most exotic
flavors in which exoplanets come, they should also reproduce their
statistical properties. This is challenging, because the predictions
of such models depend sensitively on the many key processes that come
into play. One of these key processes is the interaction of forming
planets with their parent protoplanetary disc, which is the scope of
this chapter.

Planet-disc interactions play a prominent role in the orbital
evolution of young forming planets, leading to potentially large
variations not only in their semi-major axis (a process known as
planet migration), but also in their eccentricity and
inclination. Observations (i) of hot Jupiters on orbits aligned with
the equatorial plane of their star, (ii) of systems with several
coplanar low-mass planets with short and intermediate orbital periods
(like those discovered by the Kepler mission), (iii) and of many
near-resonant multi-planet systems, are evidence that planet-disc
interactions are one major ingredient in shaping the architecture of
observed planetary systems. But, it is not the only ingredient:
planet-planet and star-planet interactions are also needed to account
for the diversity of exoplanetary systems. The long-term evolution of
planetary systems after the dispersal of their protoplanetary disc is
reviewed in the chapter by Davies et al.

This chapter commences with a general description of planet-disc
interactions in section~\ref{sec:THEORY}. Basic processes and recent
progress are presented with the aim of letting non-experts pick up a
physical intuition and a sense of the effects in planet-disc
interactions. Section~\ref{sec:APP} continues with a discussion on the
role played by planet-disc interactions in the properties and
architecture of observed planetary systems. Summary points follow in
section~\ref{sec:summary}.

\medskip
\section{\textbf{THEORY OF PLANET-DISC INTERACTIONS}}
\label{sec:THEORY}

\subsection{Orbital evolution of low-mass planets: type I migration}
\label{sec:type1}
\noindent Embedded planets interact with the surrounding disc mainly
through gravity. Disc material, in orbit around the central star,
feels a gravitational perturbation caused by the planet that can lead
to an exchange of energy and angular momentum between planet and
disc. In this section, we assume that the perturbations due to the
planet are small, so that the disc structure does not change
significantly due to the planet, and that migration is slow, so that
any effects of the radial movement of the planet can be neglected. We
will return to these issues in sections~\ref{sec:type2}
and~\ref{sec:type3}.  If these assumptions are valid, we are in the
regime of type I migration, and we are dealing with low-mass planets
(typically up to Neptune's mass).

\begin{figure}[!t]
 \epsscale{0.90}
 \plotone{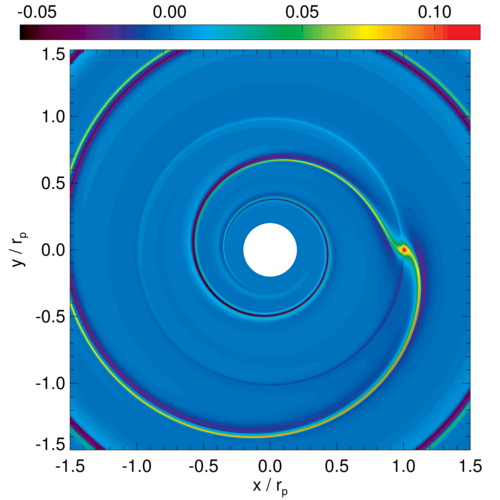}
 \epsscale{0.75}
 \plotone{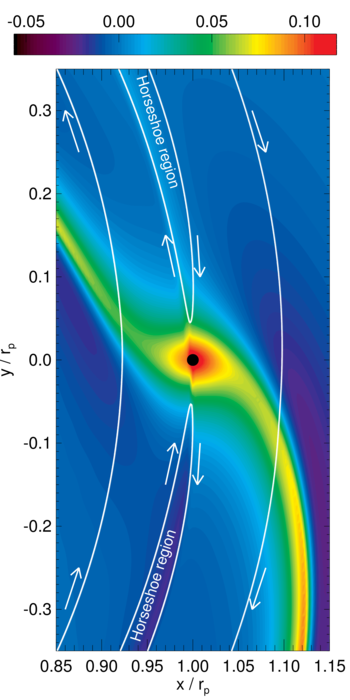}
 \caption{\small Relative perturbation of the surface density of a gaseous protoplanetary disc perturbed 
 by a 5 Earth-mass planet located at $x=r_{\rm p}$ and $y=0$. The planet induces  
 a one-armed spiral density wave -- the {\it wake} -- that propagates throughout the disc, and density 
 perturbations confined in the planet's horseshoe region. Typical gas trajectories 
 relative to the planet are shown with white curves and arrows in the 
 bottom panel.
  \label{fig:overview}
}
\end{figure}

The perturbations in the disc induced by the planet are traditionally
split into two parts: (i) a wave part, where the disc response comes
in the form of spiral density waves propagating throughout the disc
from the planet location, and (ii) a part confined in a narrow region
around the planet's orbital radius, called the planet's horseshoe
region, where disc material executes horseshoe turns relative to the
planet. An illustration of both perturbations is given in
Fig.~\ref{fig:overview}. We will deal with each of them separately
below. For simplicity, we will focus on a two-dimensional disc,
characterised by vertically averaged quantities such as the surface
density $\Sigma$. We make use of cylindrical polar coordinates
$(r,\varphi)$ centred on the star.

\subsubsection{Wave torque}
\label{sec:Lindblad}
\noindent It has been known for a long time that a planet exerting a
gravitational force on its parent disc can launch waves in the disc at
Lindblad resonances \citep{GT79, GT80}. These correspond to locations
where the gas azimuthal velocity relative to the planet matches the
phase velocity of acoustic waves in the azimuthal direction. This
phase velocity depends on the azimuthal wavenumber, the sound speed
and the epicyclic frequency, that is the oscillation frequency of a
particle in the disc subject to a small radial displacement
\citep[e.g.,][]{W97}.  At large azimuthal wavenumber, the phase
velocity tends to the sound speed, and Lindblad resonances therefore
pile up at $r=a_{\rm p}\pm 2H/3$, where $a_{\rm p}$ is the semi-major
axis of the planet and $H\ll r$ is the pressure scaleheight of the
disc \citep{Art93}. The superposition of the waves launched at
Lindblad resonances gives rise to a one-armed spiral density wave
\citep{2002MNRAS.330..950O}, called the wake (see
Fig.~\ref{fig:overview}).

It is possible to solve the wave excitation problem in the linear
approximation and calculate analytically the resulting torque exerted
on the disc using the WKB approximation \citep{GT79, LP79}.  Progress
beyond analytical calculations for planets on circular orbits has been
made by solving the linear perturbation equations numerically, as done
in 2D in \cite{KP93}. The three-dimensional case was tackled in
\cite{TTW02}, which resulted in a widely used torque formula valid for
isothermal discs only.  It is important to note that 2D calculations
only give comparable results to more realistic 3D calculations if the
gravitational potential of the planet is softened
appropriately. Typically, the softening length has to be a sizeable
fraction of the disc scaleheight \citep{MKM12}. Using this 2D softened
gravity approach, \cite{PBCK10} found that the dependence of the wave
torque\footnote{The wave torque exerted on the planet is also commonly
  referred to as the Lindblad torque.}  ($\Gamma_{\rm L}$) on disc
gradients in a non-isothermal, adiabatic disc is
\begin{equation}
\gamma\Gamma_{\rm L}/\Gamma_0 = -2.5-1.7\beta + 0.1\alpha,
\label{eqTL}
\end{equation}
where $\gamma$ is the ratio of specific heats, $\alpha$ and $\beta$
are negatives of the (local) power law exponents of the surface
density $\Sigma$ and temperature $T$ ($\Sigma \propto r^{-\alpha}$, $T
\propto r^{-\beta}$), and the torque is normalised by
\begin{equation}
\Gamma_0=\frac{q^2}{h^2}\Sigma_{\rm p}r_{\rm p}^4\Omega_{\rm p}^2,
\label{eqG0}
\end{equation}
where $q$ is the planet-to-star mass ratio, $h=H/r$ is the aspect
ratio and quantities with subscript $p$ refer to the location of the
planet. Note that in general we expect $\alpha,\beta>0$, i.e. both
surface density and temperature decrease outward. For reasonable
values of $\alpha$, the wave torque on the planet is negative: it
decreases the orbital angular momentum of the planet, and thus its
semi-major axis (the planet being on a circular orbit), leading to
inward migration. The linear approximation remains valid as long as $q
\ll h^3$ \citep{KP96}. For a disc around a Solar mass star with
$h=0.05$ this means that the planet mass needs to be much smaller than
$40$ $M_\Earth$.

The factor $\gamma$ in Eq.~(\ref{eqTL}) is due to the difference
in sound speed between isothermal and adiabatic discs
\citep{BM08a}. For discs that can cool efficiently, we expect the
isothermal result to be valid ($\gamma\rightarrow 1$), while for discs
that can not cool efficiently, the adiabatic result should hold. It is
possible to define an effective $\gamma$ that depends on the thermal
diffusion coefficient so that the isothermal regime can be connected
smoothly to the adiabatic regime \citep{PBK11}.

A generalized expression for the Lindblad torque has been derived by
\citet{Masset11} for 2D discs where the density and temperature
profiles near the planet are not power laws, like at opacity
transitions or near cavities. This generalized expression agrees well
with Eq.~(\ref{eqTL}) for power-law discs. We stress that there is to
date no general expression for the wave torque in 3D non-isothermal
discs. The analytics is involved \citep{TTW02, DAL10} and it is
difficult to measure the wave torque independently from the corotation
torque in 3D numerical simulations of planet-disc interactions.

The above discussion neglected possible effects of
self-gravity. \cite{PH05} showed that in a self-gravitating disc,
Lindblad resonances get shifted towards the planet, thereby making the
wave torque stronger. This was confirmed numerically by
\cite{BM08b}. The impact of a magnetic field in the disc and of
possibly related MHD turbulence will be considered in
Section~\ref{sec:turb}.

The normalisation factor $\Gamma_0$ sets a time scale for Type I
migration of planets on circular orbits:
\begin{equation}
\tau_0 = \frac{r_{\rm p}}{|dr_{\rm p}/dt|} = \frac12 \frac{h^2}{q}
\frac{M_{\star}}{\Sigma_{\rm p}r_{\rm p}^2}\Omega_{\rm p}^{-1},
\label{eq:tau0}
\end{equation}
where $M_{\star}$ denotes the mass of the central star.  Assuming a
typical gas surface density of $2000 \,(r_{\rm p}/1\,{\rm
  AU})^{-3/2}\,{\rm g}\,{\rm cm}^{-2}$, $M_{\star} = M_\odot$, and
$h=0.05$, the migration time scale in years at 1 Astronomical Unit
(AU) is given approximately by $1/q$. This means that an Earth-mass
planet at 1 AU would migrate inward on a time scale of $\sim
3\times10^5$ years, while the time scale for Neptune would only be
$\sim 2\times10^4$ years. All these time scales are shorter than the
expected disc life time of $10^6-10^7$ years, making this type of
migration far too efficient for planets to survive on orbits of
several AU. A lot of work has been done recently on how to stop Type I
migration or make it less efficient (see sections \ref{sec:coro} and
\ref{sec:coro_satu}).

\subsubsection{Corotation torque}
\label{sec:coro}
\noindent Most progress since PPV \citep{PPV} has been made in
understanding the torque due to disc material that on average
corotates with the planet, the corotation torque. It is possible, by
solving the linearised disc equations in the same way as for the wave
torque, to obtain a numerical value for the corotation torque. One can
show that in the case of an isothermal disc, this torque scales with
the local gradient in specific vorticity or
vortensity\footnote{Vorticity is defined here as the vertical
  component of the curl of the gas velocity. In a 2D disc model,
  specific vorticity, or vortensity, refer to vorticity divided by
  surface density.}  \citep{GT79}.  It therefore has a stronger
dependence on background surface density gradients than the wave
torque, with shallower profiles giving rise to a more positive
torque. It was nevertheless found in \cite{TTW02} that, except for
extreme surface density profiles, the wave torque always dominates
over the linear corotation torque
($\Gamma_\mathrm{c,lin}$). \cite{PBCK10} obtained, in the 2D softened
gravity approach, for a non-isothermal disc
\begin{equation}
  \gamma\Gamma_\mathrm{c,lin}/\Gamma_0 = 0.7\left(\frac{3}{2}-\alpha-\frac{2\xi}{\gamma}\right) + 2.2\xi,
\label{eqTclin}
\end{equation}
where $\xi=\beta-(\gamma-1)\alpha$ is the negative of the (local)
power law exponent of the specific entropy.  For an isothermal disc,
$\xi=0$ and the corotation torque is proportional to the vortensity
gradient.

An alternative expression for the corotation torque was derived by
\cite{W91} by considering disc material on horseshoe orbits relative
to the planet. This disc material defines the planet's horseshoe
region (see bottom panel in Fig.~\ref{fig:overview}). The torque on
the planet due to disc material executing horseshoe turns, the
horseshoe drag, scales with the vortensity gradient in an isothermal
disc, just like the linear corotation torque. It comes about because
material in an isothermal inviscid fluid conserves its
vortensity. When executing a horseshoe turn, which takes a fluid
element to a region of different vorticity because of the Keplerian
shear, conservation of vortensity dictates that the surface density
should change \citep{W91}. In a gas disc, this change in surface
density is smoothed out by evanescent pressure waves
\citep{CM09}. Nevertheless, this change in surface density results in
a torque being applied on the planet.

For low-mass planets, for which $q\ll h^3$, the half-width of the
horseshoe region, $x_{\rm s}$, is \citep{MDK06,PP09a}
\begin{equation}
  x_{\rm s} \approx 1.2 r_{\rm p} \sqrt{q/h}.
\label{eq_xs}
\end{equation}
Thus it is only a fraction of the local disc thickness. The horseshoe
drag, which scales as $x_{\rm s}^4$ \citep{W91}, therefore has the
same dependence on $q$ and $h$ as the wave torque and the linear
corotation torque.  The numerical coefficient in front is generally
much larger than for the linear corotation torque, however
\citep{PP09b,PBCK10}. Since both approaches aim at describing the same
thing, \emph{the} corotation torque, it was long unclear which result
to use. It was shown in \cite{PP09b} that whenever horseshoe turns
occur, the linear corotation torque gets replaced by the horseshoe
drag, unless a sufficiently strong viscosity is applied. It should be
noted that horseshoe turns do not exist within linear theory, making
linear theory essentially invalid for low-mass planets as far as the
corotation torque is concerned.

The corotation torque in the form of horseshoe drag is very sensitive
to the disc's viscosity and thermal properties near the
planet. \cite{PM06} found that in 3D radiation hydrodynamical
simulations, migration can in fact be directed \emph{outwards} due to
a strong positive corotation torque counterbalancing the negative wave
torque over the short duration (15 planet orbits) of their
calculations.  This was subsequently interpreted as being due to a
radial gradient in disc specific entropy, which gives rise to a new
horseshoe drag component due to conservation of entropy in an
adiabatic disc \citep{BM08a,PM08,PP08}. The situation is, however, not
as simple as for the isothermal case. It turns out that the
entropy-related horseshoe drag arises from the production of vorticity
along the downstream separatrices of the planet's horseshoe region
\citep{MC09}.  Conservation of entropy during a horseshoe turn leads
to a jump in entropy along the separatrices whenever there is a radial
gradient of entropy in the disc. This jump in entropy acts as a source
of vorticity, which in turn leads to a torque on the
planet. Crucially, the amount of vorticity produced depends on the
streamline topology, in particular the distance of the stagnation
point to the planet. An analytical model for an adiabatic disc where
the background temperature is constant was developed in \cite{MC09},
while \cite{PBCK10} used a combination of physical arguments and
numerical results to obtain the following expression for the horseshoe
drag:
\begin{equation}
  \gamma\Gamma_\mathrm{c,HS}/\Gamma_0 = 1.1\left(\frac{3}{2}-\alpha\right) + 7.9\frac{\xi}{\gamma},
\label{eqTcHS}
\end{equation}
where the first term on the right hand side is the vortensity-related
part of the horseshoe drag, and the second term is the entropy-related
part. The model derived in \cite{MC09}, under the same assumptions for
the stagnation point, yields a numerical coefficient for the
entropy-related part of the horseshoe drag of $7.0$ instead of $7.9$.

Comparing the linear corotation torque to the non-linear horseshoe
drag -- see Eqs.~(\ref{eqTclin}) and~(\ref{eqTcHS}) -- we see that
both depend on surface density and entropy gradients, but also that
the horseshoe drag is always stronger. In the inner regions of discs
primarily heated by viscous heating, the entropy profile should
decrease outward ($\xi >0$). The corotation torque should then be
positive, promoting outward migration.

The results presented above were for adiabatic discs, while the
isothermal result can be recovered by setting $\gamma=1$ and $\beta=0$
(which makes $\xi=0$ as well). Real discs are neither isothermal nor
adiabatic. When the disc can cool efficiently, which happens in the
optically thin outer parts, the isothermal result is expected to be
valid \citep[or rather the \emph{locally} isothermal result: a disc
with a fixed temperature profile, which behaves slightly differently
from a truly isothermal disc; see][]{CM09}. In the optically thick
inner parts of the disc, cooling is not efficient and the adiabatic
result should hold. An interpolation between the two regimes was
presented in \cite{PBK11} and \cite{MC10}.

\subsubsection{Saturation of the corotation torque}
\label{sec:coro_satu}
\par\noindent 
While density waves transport angular momentum away from the planet,
the horseshoe region only has a finite amount of angular momentum to
play with since there are no waves involved. Sustaining the corotation
torque therefore requires a flow of angular momentum into the
horseshoe region: unlike the wave torque, the corotation torque is
prone to {\it saturation}. In simulations of disc-planet interactions,
sustaining (or unsaturating) the corotation torque is usually
established by including a Navier-Stokes viscosity. In a real disc,
angular momentum transport is likely due to turbulence arising from
the magneto-rotational instability (MRI; see chapter by Turner et
al.), and simulations of turbulent discs give comparable results to
viscous discs \citep{BL10,BFNM11,PBH12} as far as saturation is
concerned (see also Section~\ref{sec:turb}).  The main result for
viscous discs is that the viscous diffusion time scale across the
horseshoe region has to be shorter than the libration time scale in
order for the horseshoe drag to be unsaturated \citep{M01,M02}. This
also holds for non-isothermal discs, and expressions for the
corotation torque have been derived in 2D disc models for general
levels of viscosity and thermal diffusion or cooling
\citep{MC10,PBK11}. Results from 2D radiation hydrodynamic
simulations, where the disc is self-consistently heated by viscous
dissipation and cooled by radiative losses, confirm this picture
\citep{KC08}.

The torque expressions of \citet{MC10} and \citet{PBK11} were derived
using 2D disc models, and 3D disc models are still required to get
definitive, accurate predictions for the wave and the corotation
torques. Still, the predictions of the aforementioned torque
expressions are in decent agreement with the results of 3D simulations
of planet-disc interactions. The simulations of \citet{DAL10} explored
the dependence of the total torque with density and temperature
gradients in 3D locally isothermal discs. They found the total
normalized torque $\Gamma_{\rm tot}/\Gamma_0 \approx -1.4 - 0.4\beta -
0.6\alpha$. For the planet mass and disc viscosity of these authors,
the corotation torque reduces to the linear corotation torque. Summing
Eqs.~(\ref{eqTL}) and~(\ref{eqTclin}) with $\gamma=1$ and $\xi=\beta$
yields $\Gamma_{\rm tot}/\Gamma_0 = -1.4 - 0.9\beta - 0.6\alpha$,
which is in good agreement with the results of \citet{DAL10}, except
for the coefficient in front of the temperature gradient.  The
simulations of \citet{KBK09}, \citet{AB11} and
\citet{Bitsch11_formulae} were for non-isothermal radiative discs.
\citet{AB11} considered various temperature power-law profiles and
showed that the total torque does not always exhibit a linear
dependence with temperature gradient. \citet{Bitsch11_formulae}
highlighted substantial discrepancies between the torque predictions
of \citet{MC10} and \citet{PBK11}. These discrepancies originate from
a larger half-width ($x_{\rm s}$) of the planet's horseshoe region
adopted in \citet{MC10}, which was suggested as a proxy to assess the
corotation torque in 3D. Adopting the same, standard value for $x_{\rm
  s}$ given by Eq.~(\ref{eq_xs}), one can show that the total torques
of \citet{MC10} and \citet{PBK11} are actually in very good
agreement. Both are less positive than in the numerical results of
\citet{Bitsch11_formulae}. Discrepancies originate from inherent
torque differences between 2D and 3D disc models \citep[see,
e.g.,][]{KBK09}, and possibly from a non-linear boost of the positive
corotation torque in the simulations of \citet{Bitsch11_formulae}, for
which $q \gtrsim h^3$ \citep[for locally isothermal discs,
see][]{MDK06}.

\begin{figure}[!t]
 \epsscale{0.95}
 \plotone{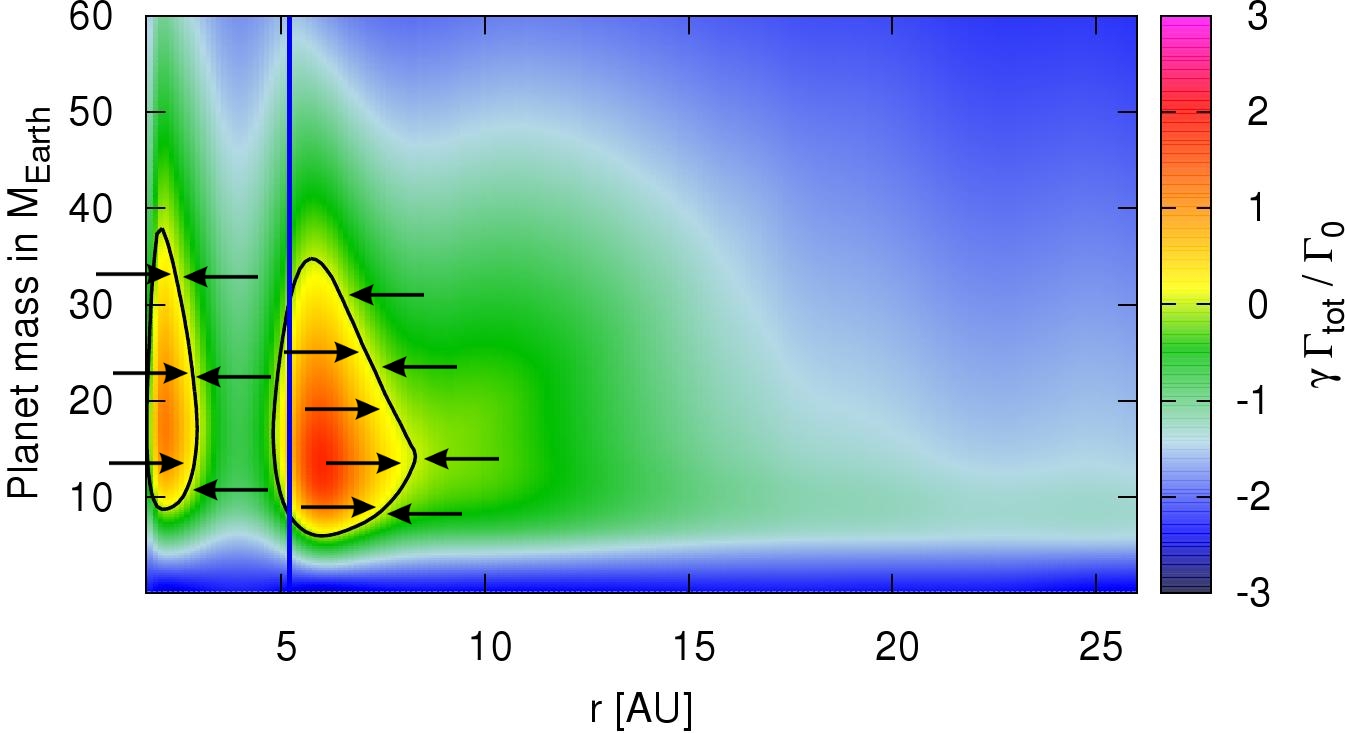}
 \plotone{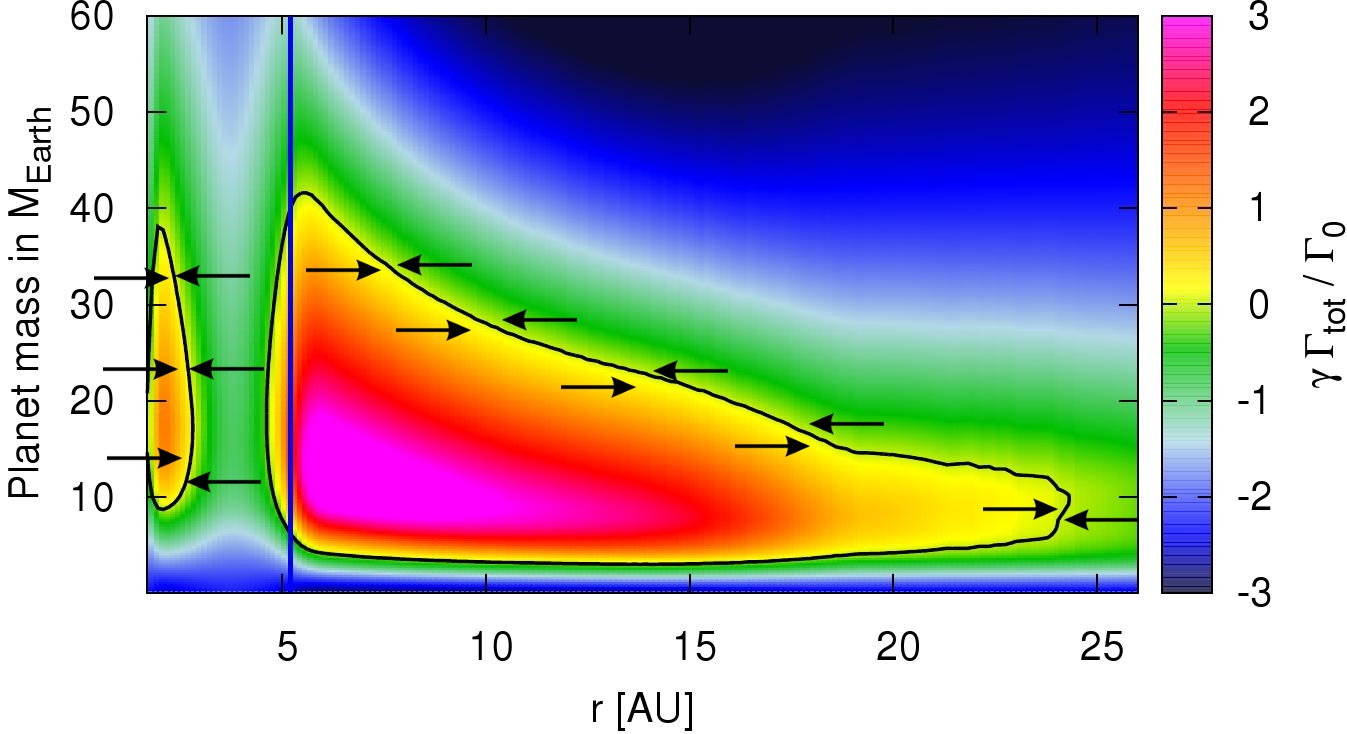}
 \caption{\small Type I migration torque on planets of different
   masses in disc models with stellar irradiation (top) and without
   (bottom). The torque expressions in \citet{PBK11} are used and 
   expressed in units of $\Gamma_0$, given by Eq.~(\ref{eqG0}).  
   Black lines delimit areas where migration is directed
   outward. Black arrows show the direction of migration, and the blue
   line the ice line location at $170$K. Adapted from
   \citet{Bitsch13}.
  \label{fig:Migration}
}
\end{figure}

The thermal structure of protoplanetary discs is determined not only
by viscous heating and radiative cooling, but also by stellar
irradiation. The effects of stellar irradiation on the disc structure
have been widely investigated using a 1+1D numerical approach
\citep[e.g.,][]{Bell97,Dullemond01}, with the goal to fit the spectral
energy distributions of observed discs. Stellar heating dominates in
the outer regions of discs, viscous heating in the inner regions. The
disc's aspect ratio should then slowly increase with increasing
distance from the star \citep{CG97}.  This increase may have important
implications for the direction and speed of type I migration which, as
we have seen above, are quite sensitive to the aspect ratio (local
value and radial profile). This is illustrated in
Fig.~\ref{fig:Migration}, which displays the torque acting on type~I
migrating planets sitting in the midplane of their disc, with and
without inclusion of stellar irradiation.  The disc structure was
calculated in \citet{Bitsch13} for standard opacities and a constant
disc viscosity (thereby fixing the density profile). Two regions of
outward migration originate from opacity transitions at the silicate
condensation line (near 0.8 AU) and at the water ice line (near 5
AU). In this model, heating by stellar irradiation becomes prominent
beyond $\approx 8$ AU. The resulting increase in the disc's aspect
ratio profile gives a shallower entropy profile ($\beta$ and thus
$\xi$ take smaller values) and therefore a smaller (though positive)
entropy-related horseshoe drag -- see Eq.~(\ref{eqTcHS}). Inclusion of
stellar irradiation therefore reduces the range of orbital separations
at which outward migration may occur \citep[see
also][]{2012ApJ...755...74K}.  The outer edge of regions of outward
migration are locations where type~I planetary cores converge, which
may lead to resonant captures \citep{Cossou13} and could enhance
planet growth if enough cores are present to overcome the trapping
process. Note in Fig.~\ref{fig:Migration} that planets up to $\sim 5
M_{\oplus}$ do not migrate outwards. This is because for such planet
masses, and for the disc viscosity in this model ($\alpha_{\nu}\sim$ a
few $\times 10^{-3}$), the corotation torque is replaced by the
(smaller) linear corotation torque.

Fig.~\ref{fig:Migration} provides a good example of how sensitive
predictions of planet migration models can be to the structure of
protoplanetary discs. Future observations of discs, for example with
ALMA, will give precious information that will help constrain
migration models.

\subsubsection{Effect of disc magnetic field and turbulence}
\label{sec:turb}
\begin{figure*}
  \epsscale{0.65} \plotone{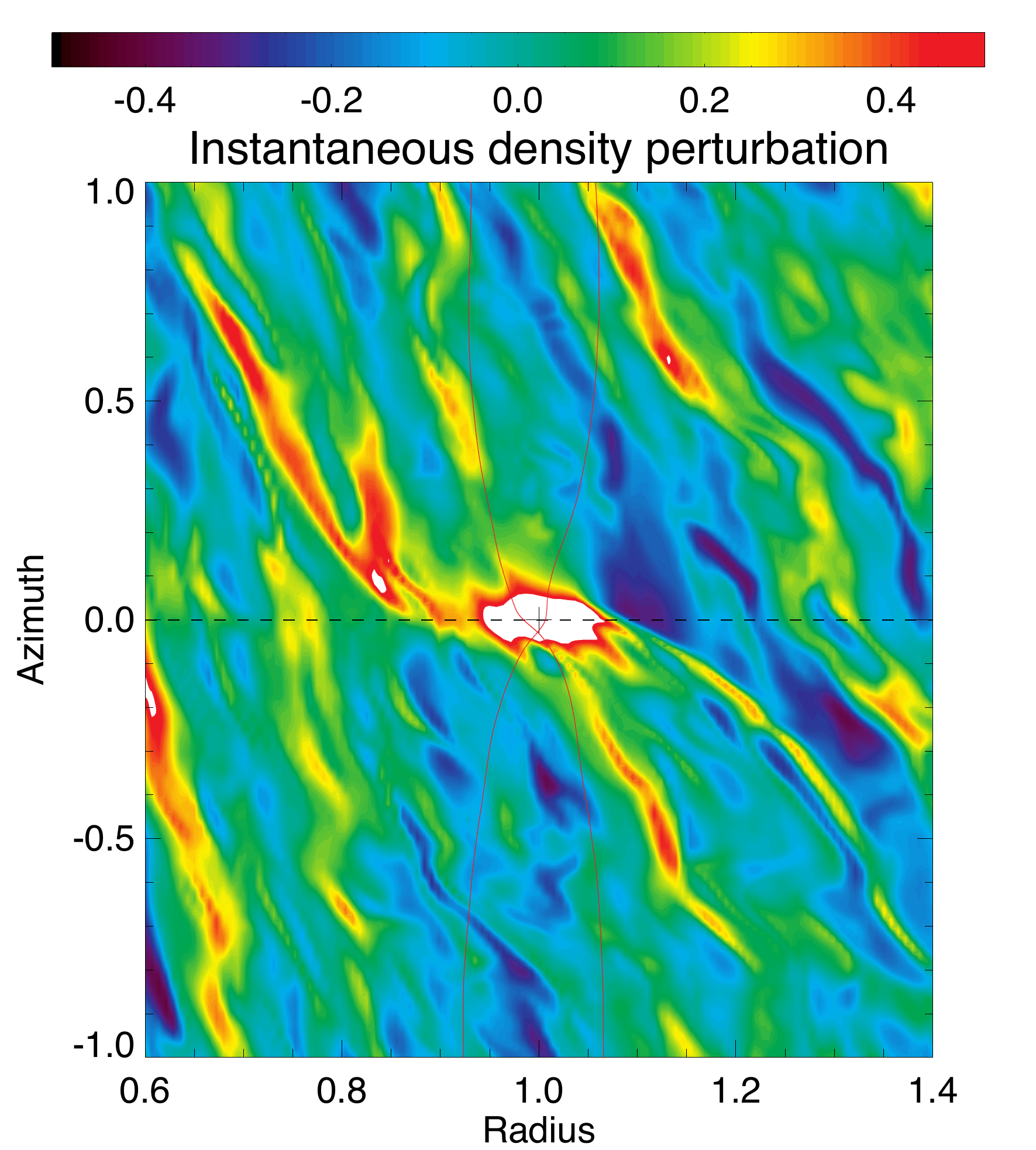}
  \plotone{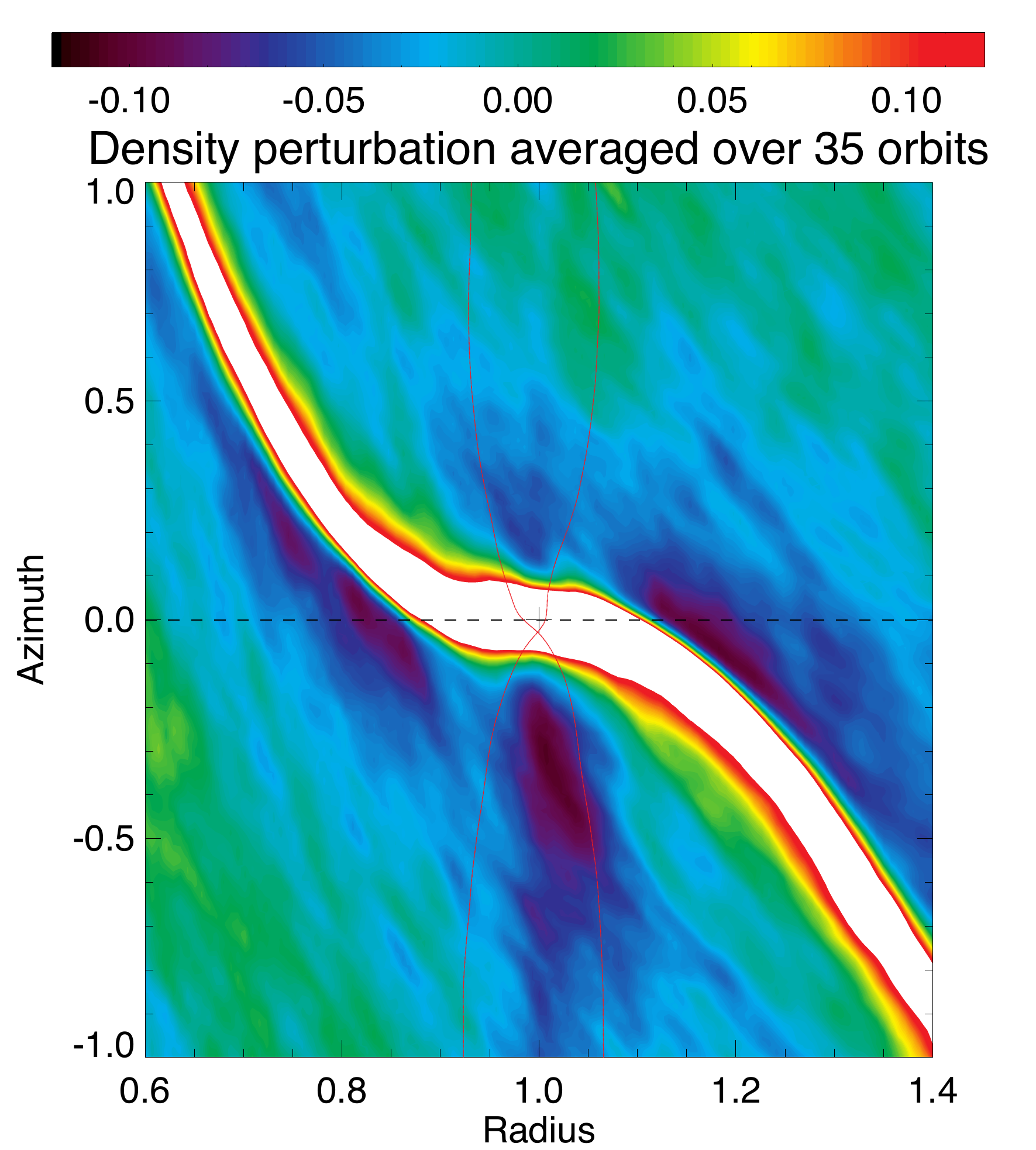} \plotone{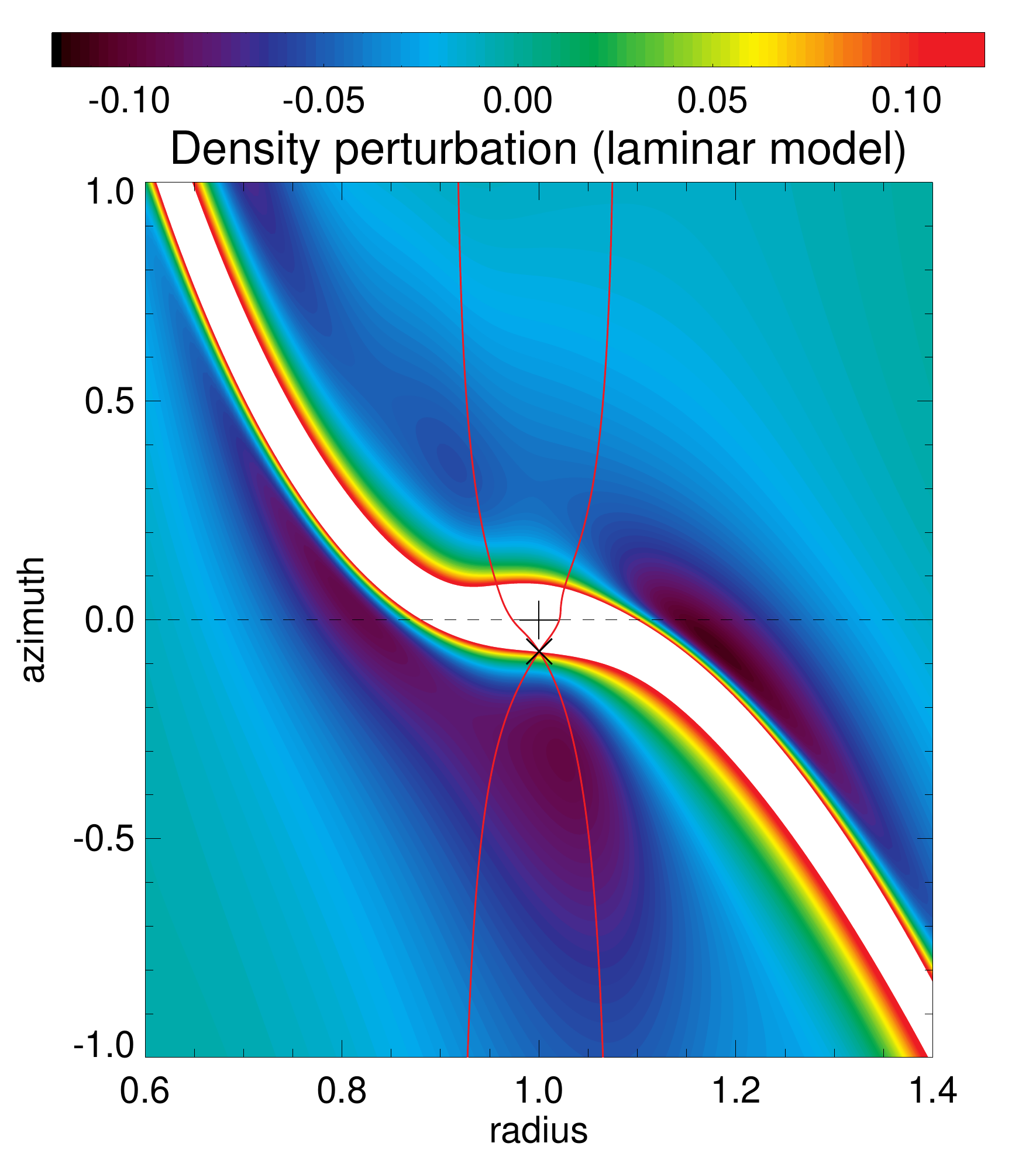}
  \caption{\small Surface density perturbation of a 3D MRI-turbulent
    disc with a planet embedded at $r=1$ and $\varphi=0$ \citep[left
    and middle panels; adapted from][]{BFNM11} and in an equivalent 2D
    laminar disc with an azimuthal magnetic field \citep[right panel,
    adapted from][]{guilet13a}.  In the turbulent simulation, the
    planet wake is barely visible in the instantaneous density
    perturbation (left), because of the large turbulent density
    fluctuations. When averaged over 35 orbits, density perturbations
    agree very well with those of the equivalent laminar disc model
    (compare the middle and right panels), revealing not only the
    planet wake, but also an asymmetric under-density confined in the
    planet's horseshoe region and arising from the azimuthal magnetic
    field. The separatrices of the planet's horseshoe region are shown
    by red curves.
   \label{fig:turbulence_magnetic}
 }
 \end{figure*}

 \noindent So far we have considered the migration of a planet in a
 purely hydrodynamical laminar disc where turbulence is modelled by an
 effective viscosity. As stressed in Section~\ref{sec:coro}, a
 turbulent viscosity is essential to unsaturate the corotation
 torque. The most likely (and best studied) source of turbulence in
 protoplanetary discs is the Magneto-Rotational Instability
 \citep[MRI;][]{balbus91} which can amplify the magnetic field and
 drive MHD turbulence by tapping energy from the Keplerian
 shear. Furthermore, the powerful jets observed to be launched from
 protoplanetary discs are thought to arise from a strong magnetic
 field \citep[likely through the magneto-centrifugal acceleration;
 see][]{ferreira06}. Magnetic field and turbulence thus play a crucial
 role in the dynamics and evolution of protoplanetary discs, and need
 to be taken into account in theories of planet-disc interactions.

 \noindent\emph{Stochastic torque driven by turbulence.} MHD turbulence
 excites non-axisymmetric density waves (see left panel in
 Fig.~\ref{fig:turbulence_magnetic}) which cause a fluctuating
 component of the torque on a planet in a turbulent disc
 \citep{NP04,nelson05,lsa04}. This torque changes sign stochastically
 with a typical correlation time of a fraction of an orbit. Because
 the density perturbations are driven by turbulence rather than the
 planet itself, the specific torque due to turbulence is independent
 of planet mass, while the (time-averaged) specific torque driving
 type I migration is proportional to the planet mass. Type I migration
 should therefore outweigh the stochastic torque for sufficiently
 massive planets and on a long-term evolution, whereas stochastic
 migration arising from turbulence should dominate the evolution of
 planetesimals and possibly small mass planetary cores
 \citep{BL10,ng10}. The stochastic torque adds a random walk component
 to planet migration, which can be represented in a statistical sense
 by a diffusion process acting on the probability distribution of
 planets \citep{jgm06,AB09}. A consequence of this is that a small
 fraction of planets may migrate to the outer parts of their disc even
 if the laminar type I migration is directed inwards.  Note that the
 presence of a dead zone around the disc's midplane, where MHD
 turbulence is quenched due the low ionization, reduces the amplitude
 of the stochastic torque \citep{oishi07,gressel11,gressel12}.

 \noindent\emph{Mean migration in a turbulent disc.} 2D hydrodynamical
 simulations of discs with stochastically forced waves have been
 carried out to mimic MHD turbulence and to study its effects on
 planet migration. Using \citet{lsa04}'s model, \citet{BL10} and
 \citet{PBH12} showed that, when averaged over a sufficiently long
 time, the torque converges toward a well-defined average value, and
 that the effects of turbulence on this average torque are well
 described by an effective turbulent viscosity and heat diffusion. In
 particular, the wave torque is little affected by turbulence, while
 the corotation torque can be unsaturated by this wake-like
 turbulence. \citet{BFNM11} and \citet{Uribe11} have shown that a
 similar conclusion holds in simulations with fully developed MHD
 turbulence arising from the MRI. \citet{BFNM11} however discovered
 the presence of an additional component of the corotation torque,
 which has been attributed to the effect of the magnetic field
 \citep[][see below]{guilet13a}. Note that the mean migration in a
 turbulent disc has been conclusively studied only for fairly massive
 planets (typically $q/h^3 \sim 0.3$) as less massive planets need a
 better resolution and a longer averaging time. It is an open question
 whether the migration of smaller planets is affected by turbulence in
 a similar way as by a diffusion process. One may wonder indeed
 whether the diffusion approximation remains valid when the width of
 the planet's horseshoe region is a small fraction of the disc
 scaleheight and therefore of the typical correlation length of
 turbulence.

 \noindent\emph{Wave torque with a magnetic field.} In addition to driving
 turbulence, the magnetic field has a direct effect on planet
 migration by modifying the response of the gas to the planet's
 potential. In particular, waves propagation is modified by the
 magnetic field and three types of waves exist: fast and slow
 magneto-sonic waves as well as Alfv\' en waves. \cite{terquem03}
 showed that for a strong azimuthal magnetic field, slow MHD waves are
 launched at the so-called magnetic resonances, which are located
 where the gas azimuthal velocity relative to the planet matches the
 phase velocity of slow MHD waves. The angular momentum carried by the
 slow MHD waves gives rise to a new component of the torque. If the
 magnetic field strength is steeply decreasing outwards, this new
 torque is positive and may lead to outward migration
 \citep{terquem03,fromang05}. A vertical magnetic field also impacts
 the resonances \citep{muto08} but its effect on the total torque
 remains to be established. In the inner parts of protoplanetary
 discs, the presence of a strong vertical magnetic field is needed to
 explain the launching of observed jets. A better understanding of the
 strength and evolution of such a vertical field
 \citep{guilet12,guilet13b} and of its effect on planet migration will
 improve the description of planet migration near the central star.

 \noindent\emph{Corotation torque with a magnetic field.} A strong
 azimuthal magnetic field can prevent horseshoe motions so that the
 corotation torque is replaced by the torque arising from magnetic
 resonances as discussed in the previous paragraph. \citet{guilet13a}
 showed that horseshoe motions take place and suppress magnetic
 resonances for weak enough magnetic fields, when the Alfv\'en speed
 is less than the shear velocity at the separatrices of the planet's
 horseshoe region. Using 2D laminar simulations with effective
 viscosity and resistivity, these authors showed that advection of the
 azimuthal magnetic field along horseshoe trajectories leads to an
 accumulation of magnetic field near the downstream separatrices of
 the horseshoe region. This accumulation in turns leads to an
 under-density at the same location to ensure approximate pressure
 balance (see right panel in Fig.~\ref{fig:turbulence_magnetic}). The
 results of these laminar simulations agree very well with those of
 the MHD turbulent simulations of \citet{BFNM11}. A rear-front
 asymmetry in the magnetic field accumulation inside the horseshoe
 region gives rise to a new component of the corotation torque, which
 may cause outward migration even if the magnetic pressure is less
 than one percent of the thermal pressure \citep{guilet13a}. This new
 magnetic corotation torque could take over the entropy-related
 corotation torque to sustain outward migration in the radiatively
 efficient outer parts of protoplanetary discs. Future studies should
 address the behaviour of the corotation torque in the dead zone, and
 in regions of the disc threaded by a vertical magnetic field. Also,
 other non-ideal MHD effects, such as the Hall effect and ambipolar
 diffusion, can have a significant impact on the MRI turbulence and on
 the disc structure
 \citep[e.g.,][]{BaiStone11,KunzLesur13,Simon13}. These non-ideal MHD
 effects still need to be explored in the context of planet-disc
 interactions.

\subsubsection{Evolution of eccentric or inclined low-mass planets}
\label{sec:ei}
\noindent We have so far considered planet-disc interactions for
low-mass planets on circular orbits. Interaction between two or more
planets migrating in their parent disc may increase eccentricities
($e$) and inclinations ($i$) -- see section~\ref{sec:multi}. We
examine below the orbital evolution of protoplanets on eccentric or
inclined orbits due to planet-disc interactions.

In the limit of small eccentricities, it can be shown that the effect
of the disc is to damp the eccentricity of type~I migrating planets
\citep[][]{GT80, Art93a, Masset08}. Similar arguments can be made for
inclined low-mass planets, for which planet-disc interactions damp the
inclination in time \citep{Tanaka04}.  \citet{PL00} have provided an
analytic expression for the eccentricity damping time scale in 2D,
while \citet{Tanaka04} derived expressions for the eccentricity and
inclination damping time scales in 3D:
\begin{equation}
  \tau_e = \frac{e}{|de/dt|} \approx 2.6\,h^2\,\tau_0,\;\;
  \tau_i = \frac{i}{|di/dt|} \approx 3.7\,h^2\,\tau_0, 
\end{equation}
where $\tau_0$ is given by Eq.~(\ref{eq:tau0}). Note that, $h^2$ being
very small, damping of $e$ and $i$ is much faster than migration. A
single low-mass planet should therefore migrate on a circular and
coplanar orbit.  The above results were confirmed by hydrodynamical
simulations \citep{Cress07,Bitsch10,Bitsch11}.  Eccentricity and
inclination damping rates can be alternatively derived using a
dynamical friction formalism \citep{Muto11,Rein12}.

We have stressed above that the migration of low-mass planets can be
directed outwards if the disc material in the horseshoe region exerts
a strong positive corotation torque on the planet. Numerical
simulations by \citet{Bitsch10} have shown that for small
eccentricities, the magnitude of the corotation torque decreases with
increasing eccentricity, which restricts the possibility of outward
migration to planets with eccentricities below a few percent. A
consequence of this restriction is to shift regions of convergent
migration to smaller radii for mildly eccentric low-mass planets
\citep[see last two paragraphs in section~\ref{sec:coro},
and][]{Cossou13}.  In a very recent study, \citet{Fendyke13} have
explored in detail the influence of orbital eccentricity on the
corotation torque for a range of disc and planet parameters.  Their
study indicates that the reason why the corotation torque decreases
with increasing $e$ is because the width of the horseshoe region
narrows as $e$ increases. Furthermore, by fitting the results from
their suite of simulations with an analytic function, they find that
the corotation torque scales with eccentricity according to the
expression $\Gamma_{\rm c}(e) = \Gamma_{\rm c}(0) \exp{(-e/e_{\rm
    f})}$, where $\Gamma_{\rm c}(e)$ is the corotation torque at
eccentricity $e$, $\Gamma_{\rm c}(0)$ that at zero eccentricity, and
$e_{\rm f}$ is an e-folding eccentricity that scales linearly with the
disc aspect ratio at the planet's orbital radius (the expression
$e_{\rm f} = h/2 +0.01$ provides a good overall fit to the
simulations). Furthermore, the Lindblad torque becomes more positive
with increasing $e$ \citep{PL00}. This sign reversal does not
necessarily lead to outward migration as the torque on an eccentric
planet changes both the planet's semi-major axis and eccentricity
\citep[see, e.g.,][]{Masset08}.

Orbital migration also changes when a low-mass planet acquires some
inclination. The larger the inclination, the less time the planet
interacts with the dense gas near the disc midplane, the smaller the
corotation torque and the migration rate. Thus, inclined low-mass
planets can only undergo outward migration if their inclination
remains below a few degrees \citep{Bitsch11}.

\subsection{Orbital evolution of massive, gap-opening planets: type II migration}
\label{sec:type2}
\subsubsection{Gap opening}
\label{sec:gapopening}
\noindent 
The wave torque described in section~\ref{sec:Lindblad} is the sum of
a positive torque exerted on the planet by its inner wake, and a
negative torque exerted by its outer wake. Equivalently, the planet
gives angular momentum to the outer disc (the disc beyond the planet's
orbital radius), and it takes some from the inner disc. If the torque
exerted by the planet on the disc is larger in absolute value than the
viscous torque responsible for disc spreading, an annular gap is
carved around the planet's orbit \citep{Lin-Papaloizou-1986a}. In this
simple one-dimensional picture, the gap width is the distance from the
planet where the planet torque and the viscous torque balance each
other \citep{Varniere-etal-2004}. However, \citet{Crida-etal-2006}
showed that the disc material near the planet also feels a pressure
torque that comes about because of the non-axisymmetric density
perturbations induced by the planet. In a steady state, the torques
due to pressure, viscosity and the planet balance all together, and
such condition determine the gap profile. The half-width of a
planetary gap hardly exceeds about twice the size of the planet's Hill
radius, defined as $r_{\rm H}=r_{\rm p}(q/3)^{1/3}$.  A gap should
therefore be understood as a narrow depleted annulus between an inner
disc and an outer disc. The width of the gap carved by a Jupiter-mass
planet does not exceed about $30\%$ of the star-planet orbital
separation.

Based on a semi-analytic study of the above torque balance,
\citet{Crida-etal-2006} showed that a planet opens a gap with bottom
density less than $10\%$ of the background density if the
dimensionless quantity $\mathcal{P}$ defined as
\begin{equation}
\mathcal{P}=\frac34\frac{H}{r_{\rm H}}+\frac{50}{q\mathcal{R}}
\label{eq:gap-criterion}
\end{equation}
is $\lesssim 1$. In Eq.~(\ref{eq:gap-criterion}), $\mathcal{R}={r_{\rm
    p}}^2\Omega_{\rm p}/\nu$ is the Reynolds number, $\nu$ the disc's
kinematic viscosity. Adopting the widely used alpha prescription for
the disc viscosity, $\nu=\alpha_{\nu} H^2\Omega$, the above
gap-opening criterion becomes
\begin{equation}
\frac{h}{q^{1/3}}+\frac{50\alpha_{\nu} h^2}{q} \lesssim 1.
\label{eq:gap-criterion2}
\end{equation}
This criterion is essentially confirmed by simulations of
MRI-turbulent discs, although the width and depth of the gap can be
somewhat different from an equivalent viscous disc model
\citep{Papa04,Zhu_inv}.

We point out that Eq.~(\ref{eq:gap-criterion}) with $\mathcal{P}=1$
can be solved analytically: the minimum planet-to-star mass ratio for
opening a deep gap as defined above is given by
\begin{equation}
  q_{\rm min} = \frac{100}{\mathcal{R}}
  \left[  \left(X+1\right)^{1/3}
  - \left(X-1\right)^{1/3}
  \right]^{-3},
     \label{eq:qcrit}
\end{equation}
with $X = \sqrt{1+3\mathcal{R}h^3/800}$, and where above quantities
are to be evaluated at the planet's orbital radius.  Taking a Sun-like
star, Eq.~(\ref{eq:qcrit}) shows that in the inner regions of
protoplanetary discs, where typically $h=0.05$ and
$\alpha_{\nu}\sim{\rm a\;few\;}\times 10^{-3}$, planets with a mass on
the order of that of Jupiter, or larger, will open a deep gap. In the
dead zone of a protoplanetary disc, where $\alpha_{\nu}$ can be one or
two orders of magnitude smaller, planet masses $\gtrsim 50 M_{\oplus}$
will open a gap. At larger radii, where planets could form by
gravitational instability, $h$ is probably near 0.1, $\alpha_{\nu}\sim
10^{-2}$, and only planets above 10 Jupiter masses could open a gap
(but, see section~\ref{sec:long}).

Eqs.~(\ref{eq:gap-criterion2}) and~(\ref{eq:qcrit}) give an estimate
of the minimum planet-to-star mass ratio for which a gap with density
contrast $\gtrsim 90\%$ is carved. Recent simulations by
\citet{Duffell13} indicate that similarly deep gaps could be opened
for mass ratios smaller than given by these equations in discs with
very low viscosities, such as what is expected in dead zones. Note
also that planets such that $\mathcal{P}$ is a few can open a gap with
a density contrast of few tens of percent. This may concern planets of
few Earth to Neptune masses in dead zones
\citep[e.g.,][]{Rafikov2002,Muto_inv,Dong11}.

\subsubsection{Type~II migration}
\label{sec:migtype2}
\noindent The formation of an annular gap around a planet splits the
protoplanetary disc into an inner disc and an outer disc, which both
repel the planet towards the centre of the gap. While the planet is
locked in its gap, it continues to migrate as the disc accretes onto
the star. Said differently, the planet follows the migration
trajectory imposed by the disc \citep{lp86}, and the migration
timescale is then the viscous accretion time, $\tau_{\nu}=r_{\rm
  p}^2/\nu$.  However, if the planet is much more massive than the gas
outside the gap, the planet will slow down the disc viscous
accretion. This occurs if $M_{\rm p}>4 \pi\Sigma_{\rm o}r_{\rm p}^2$,
where $\Sigma_{\rm o}$ is the surface density of the outer disc just
outside the gap.  When this occurs, the inner disc still accretes onto
the star, while the outer disc is held by the planet. This leads to
the partial (or total) depletion of the inner disc (see below). The
migration timescale, which is then set by the balance between the
viscous torque and the planet's inertia, is given by $\tau_{\nu}
\times (M_{\rm p} / 4 \pi\Sigma_{\rm o}r_{\rm p}^2)$.

The above considerations show that the timescale for type II migration
($\tau_{\rm II}$) is given by
\begin{equation}
\tau_{\rm II} = \tau_{\nu}\times\max\left(1,\frac{M_{\rm p}}{4\pi\Sigma_{\rm o}r_{\rm p}^2}\right).
\label{eq:tau2}
\end{equation}
Eq.~(\ref{eq:tau2}) applies when a planet carves a deep gap, that is
when $\mathcal{P}<1$. However, when $\mathcal{P}\gtrsim 1.5$ and the
density inside the gap exceeds about $20\%$ of the background density,
the planet and the gas are no longer decoupled.  The gas in the gap
exerts a corotation torque on the planet, which is usually
positive. Therefore, the migration of planets that marginally satisfy
the gap-opening criterion can be slower than in the standard type~II
migration. In particular, if the gap density is large enough, and
depending on the local density and temperature gradients, the
corotation torque can overcome the viscous torque and lead to outward
migration \citep{CM07}. Note that the drift of the planet relative to
the gas may lead to a positive feedback on migration (see
section~\ref{sec:type3}).

\subsubsection{Link with observations}
\label{sec:type2obs}
\noindent Recently, gap structures similar to what are predicted by
planet-disc interactions have been observed in the discs around
HD169142 \citep{Quanz-etal-2013} and TW~Hya
\citep{Debes-etal-2013}. The gap around HD169142 is located $\sim$ 50
AU from the star, and seems to be quite deep (the surface brightness
at the gap location is decreased by $\sim$10). The gap around TW~Hya
is located $\sim 80$ AU from the star, and is much shallower (the
decrease in surface brightness is only $30\%$).  If confirmed, these
would be the first observations of gaps in protoplanetary discs that
could be carved by a planet.

Cavities have been observed in several circumstellar discs in the past
few years.  Contrary to a gap, a cavity is characterised by the
absence of an inner disc.  In each of these transition discs,
observations indicate a lack of dust below some threshold radius that
extends from a few AU to few tens of AU. This lack of dust is
sometimes considered to track a lack of gas, but observational
evidence for accretion onto the star in some cases shows that the
cavities may be void of dust but not of gas. A narrow ring of hot dust
is sometimes detected in the central AU of these discs
\citep[e.g.,][]{Olofsson}, and this structure is often claimed to be
the signpost of a giant planet carving a big gap in the disc. It
should be kept in mind, however, that the gap opened by a planet is
usually much narrower than these observed depletions (see
section~\ref{sec:gapopening}), and rarely (completely) gas
proof. There is here a missing ingredient between the numerical
simulations of gas discs and observations. The outer edge of the gap
opened by a giant planet corresponds to a pressure maximum.  Dust
decoupled from the gas tends to accumulate there, and does not drift
through the gap. For typical disc densities and temperatures between 1
and 10 AU, decoupling is most efficient for dust particles of a few
centimetres to a meter \citep[e.g.,][]{OK10}.  Consequently, gaps
appear wider in the dust component than in the gas component, and the
inner disc could be void of dust even if not of gas
\citep{Fouchet-etal-2010,Zhu12_dust}.  Note that these authors find
that the smallest dust grains, which are well coupled to the gas,
should have a distribution identical to that of the gas and be present
inside the cavity. Therefore, the observation of a cavity should
depend on the size of the tracked dust, that is on the wavelength.
Interpreting observations thus requires to decouple the dynamics of
the gas and dust components. The coming years should see exciting,
high resolution observations of protoplanetary discs, with for
instance ALMA and MATISSE.

\subsubsection{Formation of a circumplanetary disc}
\label{sec:cpd}
\noindent The formation of a circumplanetary disc accompanies the
formation of a gap. The structure of a circumplanetary disc and the
gas accretion rate onto the planet have been investigated through 2D
hydrodynamical simulations \citep[e.g.,][]{Rivier-etal-2012}, 3D
hydrodynamical simulations
\citep{gda2003,Ayliffe09,Machida10,Tanigawa12}, and more recently
through 3D MHD simulations
\citep{Uribe13,Gressel13}. \citet{Gressel13} find accretion rates
$\sim0.01 M_{\oplus}\;{\rm yr}^{-1}$, in good agreement with previous
3D hydrodynamical calculations of viscous laminar discs. Also, in
agreement with previous assessment in non-magnetic disc environments,
they find that the accretion flow in the planet's Hill sphere is
intrinsically three-dimensional, and that the flow of gas toward the
planet moves mainly from high latitudes, rather than along the
mid-plane of the circumplanetary disc.

Another issue is to address how a circumplanetary disc impacts
migration. Being bound to the planet, the circumplanetary disc
migrates at the same drift rate as the planet's. Issues arise in
hydrodynamical simulations that discard the disc's self-gravity, as in
this case the wave torque can only apply to the planet, and not to its
circumplanetary material. This causes the circumplanetary disc to
artificially slow down migration, akin to a ball and chain. This issue
can be particularly important for type~III migration (see
Section~\ref{sec:type3}).  A simple workaround to this problem in
simulations of non self-gravitating discs is to exclude the
circumplanetary disc in the calculation of the torque exerted on the
planet \citep[see][]{cbkm09}. Another solution suggested by these
authors and also adopted by \citet{Peplinski1} is to imprint to the
circumplanetary disc the acceleration felt by the planet.

\subsubsection{Evolution of the eccentricity and inclination of gap-opening planets}
\label{sec:type2ei}
\noindent 
The early evolution of the Solar System in the primordial gas Solar
nebula should have led the four giant planets to be in a compact
resonant configuration, on quasi-circular and coplanar orbits
\citep{2007AJ....134.1790M, 2009ApJ...698..606C}. Their small but not
zero eccentricities and relative inclinations are supposed to have
been acquired after dispersal of the nebula, during a late global
instability in which Jupiter and Saturn crossed their 2:1 mean motion
resonance \citep{Tsiganis}. This is the so-called Nice model.

Many massive exoplanets have much higher eccentricities than the
planets in the Solar System. Also, recent measurements of the
Rossiter-McLaughlin effect have reported several hot Jupiters with
large obliquities, which indicates that massive planets could also
acquire a large inclination during their evolution.

Planet-disc interactions usually tend to damp the eccentricity and
inclination of massive gap-opening planets
\citep[e.g.,][]{Bitsch13b,XGP13}.  Expressions for the damping
timescales of eccentricity and inclination can be found in
\citet{Bitsch13b}. In particular, this would indicate that type~II
migration should only produce hot Jupiters on circular and
non-inclined orbits. There are, however, circumstances in which
planets may acquire fairly large eccentricities and obliquities while
embedded in their disk, which we summarise below.

\noindent{\it Eccentricity--}
The 3:1 mean motion resonance between a planet and the disc excites
eccentricity \citep{1991ApJ...381..259L}. Thus, if a planet carves a
gap that is wide enough for the eccentricity pumping effect of the 3:1
resonance to overcome the damping effect of all closer resonances, the
planet eccentricity will grow \citep[e.g.,][the disc eccentricity will
grow as well]{Papaloizou01}. Hydrodynamical simulations show that
planet-disc interactions can efficiently increase the eccentricity of
planets over $\sim 5-10$ Jupiter masses \citep{Papaloizou01,
  Bitsch13b,Dunhill13}. Eccentricity values up to $\approx 0.25$ have
been obtained in \citet{Papaloizou01}.

\noindent{\it Obliquity--}
Planets formed in a disk could have non-zero obliquities if the
rotation axes of the star and the disc are not aligned. Several
mechanisms causing misalignment have been proposed. One possibility is
that the protoplanetary disc had material with differing angular
momentum directions added to it at different stages of its life
\citep[e.g.,][]{BLP10}.  Alternatively, in dense stellar clusters, the
interaction with a temporary stellar companion could tilt the disc's
rotation axis \citep{Batygin12}.  However, both mechanisms should be
extended by including the interaction between the disc and the
magnetic field of the central star. This interaction might tilt the
star's rotation axis, and lead to misalignments even in discs that are
initially aligned with their star \citep{2011MNRAS.412.2790L}.

\subsection{Feedback of the coorbital dynamics on migration}
\label{sec:type3}
\noindent 
Under most circumstances, such as those presented in the previous
sections, the migration rate of a planet is provided by the value of
the disc torque, which depends on the local properties of the
underlying disc, but not on the migration rate itself. There are some
circumstances, however, in which the torque also depends on the drift
rate, in which case one has the constituting elements of a feedback
loop, with potentially important implications for migration. This, in
particular, is the case of giant or sub-giant planets embedded in
massive discs, which deplete (at least partially) their horseshoe
region.

\subsubsection{Criterion for migration to run away}
\noindent The corotation torque comes from material that executes
horseshoe U-turns relative to the planet. Most of this material is
trapped in the planet's horseshoe region.  However, if there is a net
drift of the planet with respect to the disc, material outside the
horseshoe region will execute a unique horseshoe U-turn relative to
the planet, and by doing so will exchange angular momentum with the
planet. This drift may come about because of migration, and/or because
the disc has a radial viscous drift.  The torque arising from
orbit-crossing material naturally scales with the mass flow rate
across the orbit, which depends on the relative drift of the planet
and the disc. It thus depends on the migration rate.

For the sake of definiteness, we consider hereafter a planet moving
inwards relative to the disc, but it should be kept in mind that the
processes at work here are essentially reversible, so that they can
also be applied to an outward moving planet. The picture above shows
that the corotation torque on a planet migrating inwards has three
contributions:
\begin{itemize}
\item[(i)] The contribution of the inner disc material flowing across
  the orbit. As this material gains angular momentum, it exerts a
  negative torque on the planet which scales with the drift rate. It
  therefore tends to increase the migration rate, and yields a
  positive feedback on the migration.
\item[(ii)] The contribution of the coorbital material in the planet's
  horseshoe region. It is two-fold. A first component arises from the
  material that exerts a horseshoe drag on the planet, which
  corresponds to the same horseshoe drag as if there was no drift
  between the disc and the planet (see section~\ref{sec:coro}).
\item[(iii)] Furthermore, as the material in the horseshoe region can
  be regarded as trapped in the vicinity of the planetary orbit, it
  has to move inward at the same rate as the planet. The planet then
  exerts on this material a negative torque, which scales with the
  drift rate. By the law of action-reaction, this trapped material
  yields a second, positive component of the horseshoe drag on the
  planet that scales with the drift rate. Thus, the contribution of
  the drifting trapped horseshoe material also yields a negative 
  feedback on the migration.
\end{itemize}

If the surface density profile of the disc is unaltered by the planet,
that is if the angular momentum profile of the disc is unaltered, then
contributions (i) and (iii) exactly cancel out \citep{mp03}. In that
case, the net corotation torque reduces to contribution (ii), and the
corotation torque expressions presented in section~\ref{sec:coro},
which have been derived assuming that the planet is on fixed circular
orbit, are valid regardless of the migration rate. Conversely, if the
planet depletes, at least partly, its horseshoe region (when
$\mathcal{P} \lesssim$ a few), contributions (i) and (iii) do not
cancel out, and the net corotation torque depends on the migration
rate.

The above description shows that the coorbital dynamics causes a
feedback on migration when planets open a gap around their
orbit. There are two key quantities to assess in order to determine
when the feedback causes the migration to run away. The first quantity
is called the coorbital mass deficit ($\delta M$). It represents the
mass that should be added to the planet's horseshoe region so that it
has the average surface density of the orbit-crossing flow. The second
quantity is the sum of the planet mass ($M_{\rm p}$) and of the
circumplanetary disc mass ($M_{\rm CPD}$), that is the quantity
$\tilde M_{\rm p}=M_{\rm p}+M_{\rm CPD}$. As shown in \cite{mp03}, two
regimes may occur. If $\tilde M_{\rm p}>\delta M$, the coorbital
dynamics accelerates the migration, but there is no runaway. On the
contrary, if $\tilde M_{\rm p} < \delta M$, migration runs away. A
more rigorous derivation performed by \citet{mp03} shows that the
coorbital mass deficit actually features the inverse vortensity in
place of the surface density, but the same qualitative picture holds.

\subsubsection{Properties of type~III migration}
\noindent When migration runs away, the drift rate has an
exponentially growing behaviour until the so-called fast regime is
reached, in which the planet migrates a sizable fraction of the
horseshoe width in less than a libration time. When that occurs, the
drift rate settles to a finite, large value, which defines the regime
of type~III migration. As stressed earlier, this drift can either be
outward or inward. The occurrence of type~III migration with varying
the planet mass, the disc's mass, aspect ratio, and viscosity was
discussed in detail in \citet{Masset08}.  The typical migration
timescale associated with type III migration, which depends on the
disc mass \citep{2010MNRAS.405.1473L}, is of the order of a few
horseshoe libration times. For the large planetary masses prone to
type III migration, which have wide horseshoe regions hence short
libration times, this typically amounts to a few tens of orbits.

Type~III migration can in principle be outwards. For this to happen,
an initial seed of outward drift needs to be applied to the planet 
\citep{Peplinski3,mp03,Masset08}. Nevertheless, all outward episodes
of type~III migration reported so far have been found to stall and
revert back to inward migration. Interestingly, gravitationally
unstable outer gap edges may provide a seed of outward type~III
migration and can bring massive planets to large orbital distances
\citep{2012MNRAS.421..780L,2013arXiv1306.2514C}.  An alternative
launching mechanism for outward type~III migration is the outward
migration of a resonantly locked pair of giant planets \citep[][see
also section~\ref{sec:ms01}]{massnel2001}, which is found to trigger
outwards runaways at larger time \citep{Masset08}.

The migration regime depends on how the coorbital mass deficit
compares with the mass of the planet and its circumplanetary disc. It
is thus important to describe correctly the build up of the
circumplanetary material and the effects of this mass on the dynamics
of the gas and planet.  \citet{2005MNRAS.358..316D} find, using a
nested grid code, that the mass reached by the CPD depends heavily on
the resolution for an isothermal equation of state. \citet{Peplinski1}
circumvent this problem by adopting an equation of state that not only
depends on the distance to the star, but also on the distance to the
planet, in order to prevent an artificial flood of the CPD, and to
obtain numerical convergence at high resolution. Furthermore, as we
have seen in Section~\ref{sec:cpd}, in simulations that discard
self-gravity, the torque exerted on the planet should exclude the
circumplanetary disc. \citet{2005MNRAS.358..316D} find indeed that
taking that torque into account may inhibit type~III migration.

Type~III migration has allowed to rule out a recent model of the Solar
Nebula \citep{2007ApJ...671..878D}, more compact than the standard
model of \citet{1981PThPS..70...35H}. Indeed,
\citet{2009ApJ...698..606C} has shown that Jupiter would be subject to
type~III migration in Desch's model and would not survive over the
disc's lifetime. The occurrence of type~III migration may thus provide
an upper limit to the surface density of the disc models in systems
known to harbor giant planets at sizable distances from their host
stars.

It has been pointed out above that the exact expression of the
coorbital mass deficit involves the inverse vortensity rather than the
surface density across the horseshoe region. This has some importance
in low-viscosity discs: vortensity can be regarded as materially
conserved except during the passage through the shocks triggered by a
giant planet, where vortensity is gained or destroyed. The
corresponding vortensity perturbation can be evaluated analytically
\citep{2010MNRAS.405.1473L}. Eventually, the radial vortensity
distribution around a giant planet exhibits a characteristic two-ring
structure on the edges of the gap, which is unstable
\citep{lovelace99,li2000} and prone to the formation of vortices
\citep{2010MNRAS.405.1473L}. The resulting vortensity profile
determines the occurrence of type~III migration. If vortices form at
the gap edges, the planet can undergo non-smooth migration with
episodes of type~III migration that bring the planet inwards over a
few Hill radii (a distance that is independent of the disc mass),
followed by a stalling and a rebuild of the vortensity rings
\citep{2010MNRAS.405.1473L}.

The above results have been obtained for fixed-mass planets. However,
for the high gas densities required by the type III migration regime,
rapid growth may be expected. Using 3D hydrodynamical simulations with
simple prescriptions for gas accretion, \citet{DAngelo08} find that a
planetary core undergoing rapid runaway gas accretion does not
experience type~III migration, but goes instead from the type~I to the
type~II migration regime.  Future progress will be made using more
realistic accretion rates, like those obtained with 3D
radiation-hydrodynamics calculations \citep[e.g.,][]{DAngelo13}.

\subsubsection{Feedback of coorbital dynamics on type I migration}
\noindent It has been shown that type~I migration in adiabatic discs
could feature a kind of feedback reminiscent of type~III
migration. The reason is that in adiabatic discs, the corotation
torque depends on the position of the stagnation point relative to the
planet (see section~\ref{sec:coro}). This position, in turn, depends
on the migration rate, so that there is here as well a feedback of the
coorbital dynamics on migration.  \citet{MC09} found that this
feedback on type~I migration is negative, and that is has only a
marginal impact on the drift rate for typical disc masses.

\subsection{Orbital evolution of multi-planet systems}
\label{sec:multi}
\noindent So far, we have examined the orbital evolution of a single
planet in a protoplanetary disc, while about $1/3$ of confirmed
exoplanets reside in multi-planetary systems (see exoplanets.org).  In
such systems, the gravitational interaction between planets can
significantly influence the planet orbits, leading, in particular, to
important resonant processes which we describe below. A fair number of
multi-planetary systems is known to have at least two planets in
mean-motion resonance. \citet{2012OLEB...42..113S} list for example 32
resonant or near-resonant systems, with many additional Kepler
candidate systems.  The mere existence of these resonant systems is
strong evidence that dissipative mechanisms changing planet semi-major
axes must have operated. The probability of forming resonant
configurations in situ is likely small \citep[e.g.,][]{Beauge12}.

\subsubsection{Capture in mean-motion resonance}
\label{sec:mmr}
\noindent We consider a system of two planets that undergo migration
in their disc. If the migration drift rates are such that the mutual
separation between the planets increases, i.e. when divergent
migration occurs, the effects of planet-planet interactions are small
and no resonant capture occurs. Conversely, resonant capture occurs
for convergent migration under quite general conditions, which we
discuss below.

\begin{figure}[!t]
\epsscale{0.95}
 \plotone{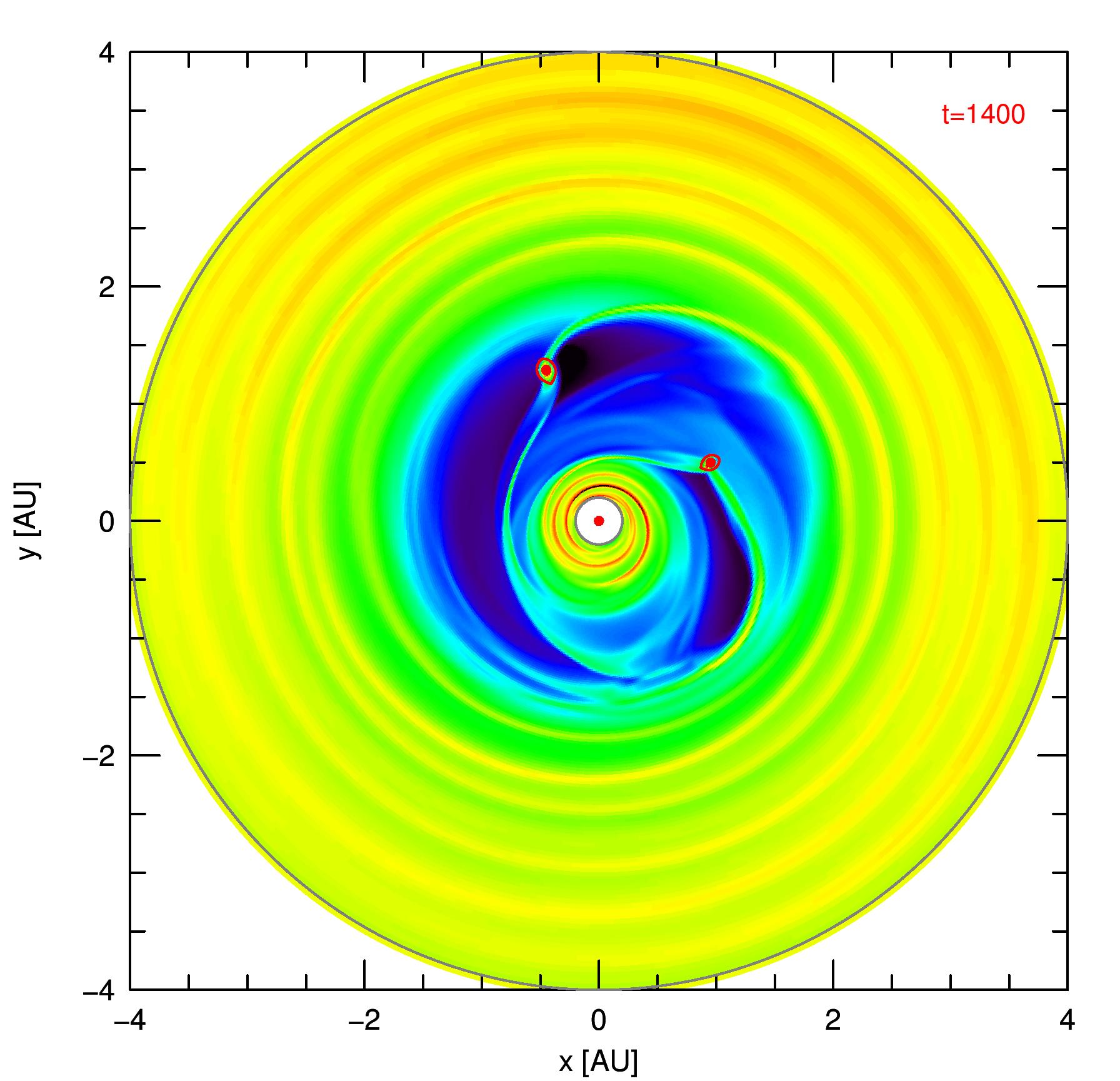}
 \caption{\small Surface density of a disc with two Jupiter-mass
   planets located in a common gap, and engaged in a 2:1 mean-motion
   resonance. The inner planet mainly interacts with the inner disc,
   and the outer planet with the outer disc, which helps to maintain
   the resonant configuration. From \citet{KN12}.}
 \label{fig:multi-planet}
\end{figure}
Planets can approach each other from widely separated orbits if they
have fairly different migration rates, or if they form in close
proximity and are sufficiently massive to carve a common gap (see
Fig.~\ref{fig:multi-planet}). In the latter case, the outer disc
pushes the outer planet inwards, the inner disc pushes the inner
planet outwards, causing convergence. If the planets approach a
commensurability, where the orbital periods are the ratio of two
integers, orbital eccentricities will be excited and resonant capture
may occur. Whether or not resonant capture occurs hinges primarily on
the time the planets take to cross the resonance. Capture requires the
convergent migration timescale to be longer than the libration
timescale associated with the resonance width
\citep{2001A&A...374.1092S}.  Otherwise, the two-planet system does
not have enough time for the resonance to be excited: the two planets
will pass through the resonance and no capture will occur
\citep[e.g.,][]{2006MNRAS.365.1367Q,2011MNRAS.413..554M}. Due to the
sensitivity of the resonant capture to the migration process, the
interpretation of observed resonant planetary systems provides
important clues about the efficiency of disc-planets interactions.

A mean-motion resonance between two planets occurs when their orbital
frequencies satisfy
\begin{equation}
\label{eq:resonance}
    ( p + q ) \Omega_2 - p \Omega_1  = 0, 
\end{equation}
where $\Omega_i$ is the angular velocity of the two planets, and where
subscripts 1 and 2 refer to quantities of the inner and outer planets,
respectively. In Eq.~(\ref{eq:resonance}), $q$ and $p$ are positive
integers, and $q$ denotes the order of the resonance. The above
condition for a mean-motion resonance can be recast in terms of the
planets semi-major axes, $a_i$, as
\begin{equation}
\label{eq:resonance-semi}
    \frac{a_2}{a_1}  = \left( \frac{p + q}{p} \right)^{2/3}.
\end{equation}
Formally, a system is said to be in a $p+q$\,:\,$p$ mean-motion
resonance if at least one of the resonant angles is librating,
i.e. has a dynamical range smaller than $2\pi$. The resonant angles
($\phi_{1,2}$) are defined as
\begin{equation}
\label{eq:ResAngles}
\phi_{1,2} = (p+q) \lambda_2 - p \lambda_1 - q \, \varpi_{1,2},
\end{equation}
where $\lambda_{i}$ denotes the planets longitude, and $\varpi_i$ the
longitude of their pericentre. The difference in pericentre longitudes
is often used to characterise resonant behaviour. For instance, when
that quantity librates, the system is said to be in apsidal
corotation.  This means that the two apsidal lines of the resonant
planets are always nearly aligned, or maintain a constant angle
between them. This is the configuration that the planets end up with
when they continue their inward migration after capture in resonance
\citep[e.g.,][]{Kley04}.

Several bodies in our Solar System are in mean-motion resonance. For
example, the Jovian satellites Io, Europa and Ganymede are engaged in
a so-called $1$:$2$:$4$ Laplace resonance, while Neptune and Pluto (as
well as the Plutinos) are in a $3$:$2$ mean-motion resonance. However,
out of the 8 planets in the Solar System, not a single pair is
presently in a mean-motion resonance. According to the Nice model for
the early Solar System, this might have been different in the past
(see section~\ref{sec:type2ei}).

The question of which resonance the system may end up in depends on
the mass, the relative migration speed, and the initial separation of
the planets \citep{2002MNRAS.333L..26N}.  Because the 2:1 resonance
($p=1, q=1$) is the first first-order resonance that two initially
well-separated planets encounter during convergent migration, it is
common for planets to become locked in that resonance, provided
convergent migration is not too rapid.

After a resonant capture, the eccentricities increase and the planets
generally migrate as a joint pair, maintaining a constant orbital
period ratio (see, however, the last two paragraphs in
section~\ref{sec:johnlow}). Continued migration in resonance drives
the eccentricities up, and they would increase to very large values in
the absence of damping agents, possibly rendering the system
unstable. The eccentricity damping rate by the disc, $\dot{e}$, is
often parametrized in terms of the migration rate $\dot{a}$ as
\begin{equation}
\label{eq:K}
\left| \frac{\dot{e}}{e} \right|= K \left| \frac{\dot{a}}{a}\right|,
\end{equation}
with $K$ a dimensionless constant. For low-mass planets, typically
below 10 to 20 $M_{\oplus}$, eccentricity damping occurs much more
rapidly than migration: $K \sim {\cal{O}}[h^{-2}] \sim$ a few hundred
in locally isothermal discs (see Section~\ref{sec:ei}) and may take
even larger values in radiative discs. High-mass planets create gaps
in their disc (see Section~\ref{sec:type2}) and the eccentricity
damping is then strongly reduced.  For the massive planets in the
GJ~876 planetary system, the three-body integrations of
\citet{2002ApJ...567..596L}, in which migration is applied to the
outer planet only, showed that $K \sim 100$ can reproduce the observed
eccentricities.  If massive planets orbit in a common gap, as in
Fig.~\ref{fig:multi-planet}, the disc parts on each side of the gap
may act as damping agents, and it was shown by the 2D hydrodynamical
simulations of \citet{Kley04} that such configuration gives $K\sim$
5--10.

Disc turbulence adds a stochastic component to convergent migration,
which may prevent resonant capture or maintenance of a resonant
configuration. This has been examined in N-body simulations with
prescribed models of disc-planet interactions and of disc turbulence
\citep[e.g.,][]{2011ApJ...726...53K, Rein12b}, and in hydrodynamical
simulations of planet-disc interactions with simplified turbulence
models \citep{PBH11,PRK13}.

\subsubsection{Application to specific multi-planet systems}
\label{sec:multiobs} 
\noindent Application of above considerations leads to excellent
agreement between theoretical evolution models of resonant capture and
the best observed systems, in particular GJ~876
\citep{2002ApJ...567..596L,2005A&A...437..727K,2008A&A...483..325C}.
Because resonant systems most probably echo an early formation via
disc-planets interactions, the present dynamical properties of
observed systems can give an indicator of evolutionary history. This
has been noticed recently in the system HD~45364, where two planets in
3:2 resonance have been discovered by \citet{2009A&A...496..521C}.
Fits to the data give semi-major axes $a_1 = 0.681$AU and
$a_2=0.897$AU, and eccentricities $e_1 = 0.168$ and $e_2=0.097$,
respectively.  Non-linear hydrodynamic simulations of disc-planets
interactions have been carried out for this system by
\citet{2010A&A...510A...4R}. For suitable disc parameters, the planets
enter the 3:2 mean-motion resonance through convergent
migration. After the planets reached their observed semi-major axis, a
theoretical RV-curve was calculated. Surprisingly, even though the
simulated eccentricities ($e_1 = 0.036, e_2 = 0.017$) differ
significantly from the data fits, the theoretical model fits the
observed data points as well as the published best fit solution
\citep{2010A&A...510A...4R}.  The pronounced dynamical differences
between the two orbital fits, which both match the existing data, can
only be resolved with more observations. Hence, HD~45364 serves as an
excellent example of a system in which a greater quantity and quality
of data will constrain theoretical models of this interacting
multi-planetary system.
 
\subsubsection{Eccentricity and inclination excitations}
\noindent Another interesting observational aspect where convergent
migration due to disc-planets interactions may have played a prominent
role is the high mean eccentricity of extrasolar planets ($\sim 0.3$).
As discussed above, for single planets, disc-planet interactions
nearly always lead to eccentricity damping, or, at best, to modest
growth for planets of few Jupiter masses (see
section~\ref{sec:type2ei}).  Strong eccentricity excitation may occur,
however, during convergent migration and resonant capture of two
planets. Convergent migration of three massive planets in a disc may
lead to close encounters that significantly enhance planet
eccentricities \citep{MBS10} and inclinations. In
the latter case, an inclination at least $\gtrsim 20-40$ degrees
between the planetary orbit and the disc may drive Kozai cycles.
Under disc-driven Kozai cycles, the eccentricity increases to large
values and undergoes damped oscillations with time in anti-phase with
the inclination \citep{Teyssandier,Bitsch13b,XGP13}.

As the disc slowly dissipates, damping will be strongly reduced. This
may leave a resonant system in an unstable configuration, triggering
dynamical instabilities \citep{2003Icar..163..290A}. Planet-planet
scattering may then pump eccentricities to much higher values. This
scenario has been proposed to explain the observed broad distribution
of exoplanet eccentricities
\citep{Chatterjee08,Juric08,2010ApJ...714..194M}. However, note that
the initial conditions taken in these studies are unlikely to result
from the evolution of planets in a protoplanetary disc \citep{Lega13}.

Using N-body simulations with 2 or 3 giant planets and prescribed
convergent migration, but no damping, \citet{LT09,LT11} found that, as
the eccentricities raise to about 0.4 due to resonant interactions,
planet inclinations could also be pumped under some conditions. The
robustness of this mechanism needs to be checked by including
eccentricity and inclination damping by the disc.

\subsubsection{Low-mass planets}
\noindent Disc-planets interactions of several protoplanets $(5-20
M_{\oplus})$ undergoing type~I migration leads to crowded systems
\citep{2008A&A...482..677C}. These authors find that protoplanets
often form resonant groups with first-order mean-motion resonances
having commensurabilities between 3:2 - 8:7 \citep[see
also][]{2005AJ....130.2884M, Papa05}.  Strong eccentricity damping
allows these systems to remain stable during their migration. In
general terms, these simulated systems are reminiscent of the low-mass
planet systems discovered by the Kepler mission, like Kepler-11
\citep{2011Natur.470...53L}.  The proximity of the planets to the star
in that system, and their near coplanarity, hints strongly toward a
scenario of planet formation and migration in a gaseous protoplanetary
disc.

\subsubsection{Possible reversal of the migration}
\label{sec:ms01}
\noindent
\citet{massnel2001} have shown that a pair of close giant planets can
migrate outwards. In this scenario, the inner planet is massive enough
to open a deep gap around its orbit and undergoes type II migration.
The outer, less massive planet opens a partial gap and migrates
inwards faster than the inner planet. If convergent migration is rapid
enough for the planets to cross the 2:1 mean-motion resonance, and to
lock in the 3:2 resonance, the planets will merge their gap and start
migrating outwards together. The inner planet being more massive than
the outer one, the (positive) torque exerted by the inner disc is
larger than the (negative) torque exerted by the outer disc, which
results in the planet pair moving outwards. Note that to maintain
joint outward migration on the long term, gas in the outer disc has to
be funnelled to the inner disc upon embarking onto horseshoe
trajectories relative to the planet pair.  Otherwise, gas would pile
up at the outer edge of the common gap, much like a snow-plough, and
the torque balance as well as the direction of migration would
ultimately reverse.

The above mechanism of joint outward migration of a resonant planet
pair relies on an asymmetric density profile across the common gap
around the two planets. This mechanism is therefore sensitive to the
disc's aspect ratio, viscosity, and to the mass ratio between the two
planets, since they all impact the density profile within and near the
common gap. If, for instance, the outer planet is too small, the disc
density beyond the inner planet's orbit will be too large to reverse
the migration of the inner planet (and, thus, of the planet
pair). Conversely, if the outer planet is too big, the torque
imbalance on each side of the common gap will favour joint inward
migration. Numerical simulations by \citep{DAngeloMarzari} showed that
joint outward migration works best when the mass ratio between the
inner and outer planets is comparable to that of Jupiter and Saturn
\citep[see also][]{2008A&A...482..333P}. In particular,
\citet{Morbidelli-Crida-2007} found that the Jupiter-Saturn pair could
avoid inward migration and stay beyond the ice line during the gas
disc phase. Their migration rate depends on the disc properties, but
could be close to stationary for standard values.

More recently, \cite{Walsh11} have proposed that Jupiter first
migrated inwards in the primordial Solar nebula down to $\sim$ 1.5 AU,
where Saturn caught it up. Near that location, after Jupiter and
Saturn have merged their gap and locked into the 3:2 mean-motion
resonance, both planets would have initiated joint outward migration
until the primordial nebula dispersed. This scenario is known as {\it
  the Grand Tack}, and seems to explain the small mass of Mars and the
distribution of the main asteroid belt.

\medskip
\section{\textbf{PLANET FORMATION AND MIGRATION: COMPARISON WITH OBSERVATIONS}}
\label{sec:APP}
\noindent The previous section has reviewed basics of, and recent
progress on planet-disc interactions. We continue in this section with
a discussion on the role played by planet-disc interactions in the
properties and architecture of observed planetary systems.
Section~\ref{sec:short} starts with planets on short-period orbits.
Emphasis is put on hot Jupiters, including those with large spin-orbit
misalignments (section~\ref{sec:johngiant}), and on the many low-mass
candidate systems uncovered by the Kepler mission
(section~\ref{sec:johnlow}). Section~\ref{sec:long} then examines how
planet-disc interactions could account for the massive planets
recently observed at large orbital separations by direct imaging
techniques. Finally, section~\ref{sect:migrat-obs} addresses how well
global models of planet formation and migration can reproduce the
statistical properties of exoplanets.

\subsection{Planets on short-period orbits}
\label{sec:short}

\subsubsection{Giant planets}
\label{sec:johngiant}
\noindent The discovery of 51 Pegasi b \citep{MarcyButler95} on a
close-in orbit of $4.2$ days as the first example of a hot Jupiter led
to the general view that giant planets, which are believed to have
formed beyond the ice line at $\gtrsim 1$ AU, must have migrated
inwards to their present locations. Possible mechanisms for this
include type II migration (see section~\ref{sec:type2}) and either
planet-planet scattering, or Kozai oscillations induced by a distant
companion leading to a highly eccentric orbit which is then
circularized as a result of tidal interaction with the central star
\citep[see, e.g.,][and references therein]{PT06,KN12,BM13}.  The
relative importance of these mechanisms is a matter of continuing
debate.

\begin{figure}[!t]
  \epsscale{1.05} \plotone{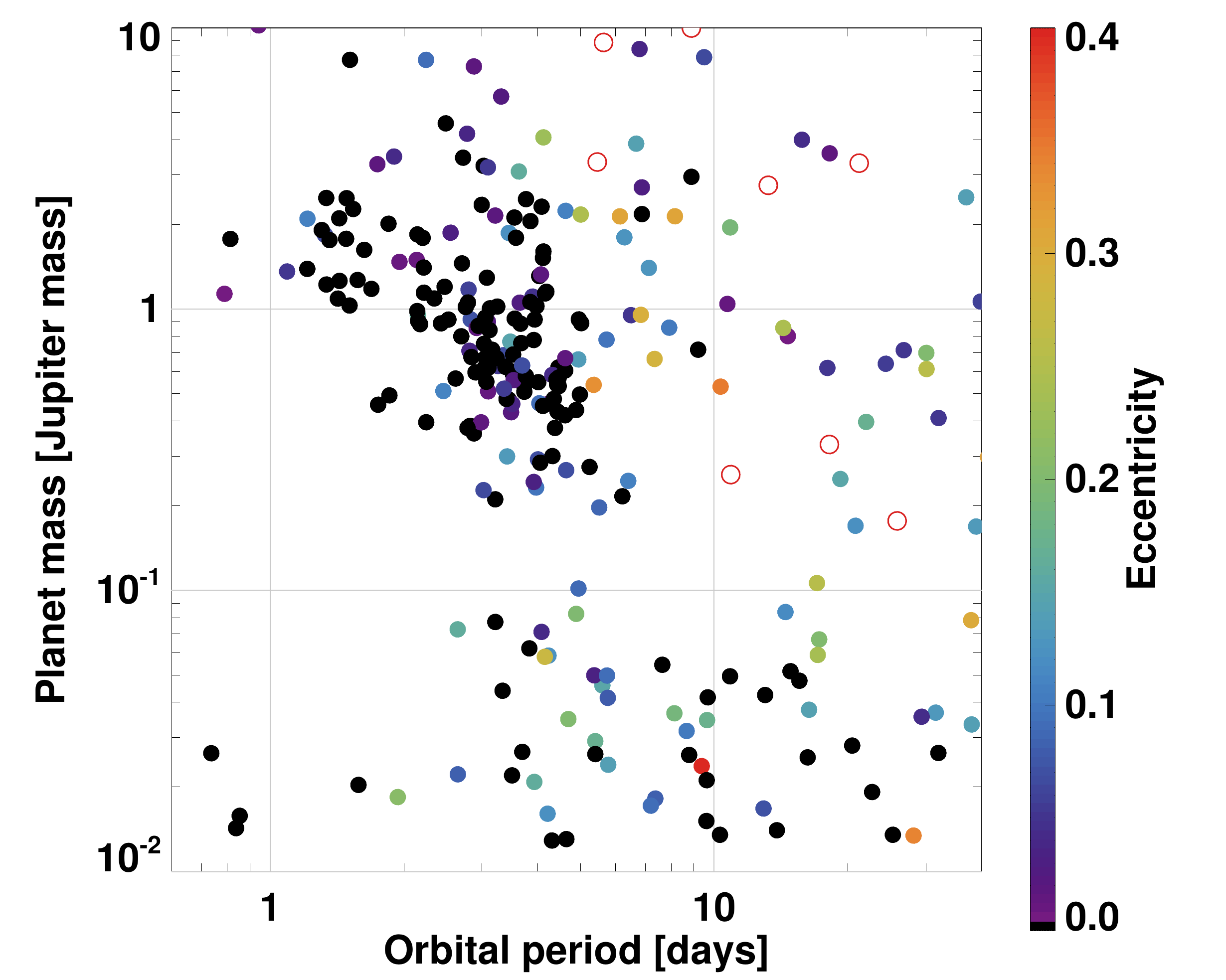}
 \caption{ \small Mass as a function of orbital period for confirmed
   exoplanets on short-period orbits. The colour of the dots
   represents the magnitude of the orbital eccentricity as indicated
   in the colour bar (white dots with a red circle correspond to
   eccentricities larger than 0.4). Data were extracted from
   exoplanets.org.
   \label{fig:jp1}}
\end{figure}
The relationship between planet mass and orbital period for confirmed
exoplanets on short-period orbits is shown in Fig.~\ref{fig:jp1}. The
hot Jupiters are seen to be clustered in circular orbits at periods in
the range $3-5$ days. For planet masses in the range $0.01 - 0.1$
Jupiter masses, there is no corresponding clustering of orbital
periods, indicating that this is indeed a feature associated with hot
Jupiters.

Measurements of the Rossiter-McLaughlin effect
\citep[e.g.,][]{Triaud10} indicate that around one third of hot
Jupiters orbit in planes that are significantly misaligned with the
equatorial plane of the central star. This is not expected from
disc-planet interactions leading to type II migration, and so has led
to the alternative mechanisms being favoured.  Thus \citet{Albrecht12}
propose that hot Jupiters are placed in an isotropic distribution
through dynamical interactions, and are then circularized by tides,
with disc-driven migration playing a negligible role. They account for
the large fraction of misaligned hot Jupiters around stars with
effective temperatures $\gtrsim 6200$~K, and the large fraction of
aligned hot Jupiters around cooler stars, as being due to a very large
increase in the effectiveness of the tidal processes causing alignment
for the cooler stars.

However, there are a number of indications that the process of hot
Jupiter formation does not work in this way, and that a more gentle
process such as disc-driven migration has operated on the
distribution. We begin by remarking the presence of significant
eccentricities for periods $\gtrsim 6$ days in Fig. \ref{fig:jp1}. The
period range over which tidal effects can circularize the orbits is
thus very limited. Giant planets on circular orbits with periods
greater than 10 days, which are interior to the ice line, must have
been placed there by a different mechanism.  For example, 55 Cnc b, a
0.8 $M_{\rm J}$ planet on a 14 day, near-circular orbit exterior to
hot super-Earth 55 Cnc e \citep{2010ApJ...722..937D} is a good case
for type II migration having operated on that system. There is no
reason to suppose that a smooth delivery of hot Jupiters through type
II migration would not function at shorter periods.

There are also issues with the effectiveness of the tidal process
\citep[see][]{RogersLin13}.  It has to align inclined circular orbits
without producing inspiral into the central star. To avoid the latter,
the components of the tidal potential that act with non-zero forcing
frequency in an inertial frame when the star does not rotate, have to
be ineffective. Instead one has to rely on components that appear to
be stationary in this limit.  These have a frequency that is a
multiple of the stellar rotation frequency, expected to be
significantly less than the orbital frequency, as viewed from the star
when it rotates.  As such components depend only on the time averaged
orbit, they are insensitive to the sense of rotation in the orbit.
Accordingly there is a symmetry between prograde and retrograde
aligning orbits with respect to the stellar equatorial plane.  This is
a strict symmetry when the angular momentum content of the star is
negligible compared to that of the orbit, otherwise there is a small
asymmetry \citep[see][]{RogersLin13}. Notably, a significant
population of retrograde orbits with aligned orbital planes, that is
expected in this scenario, is not observed.

\citet{Dawson13} have recently examined the dependence of the
relationship between mass and orbital period on the metallicity of the
central star. They find that the pile up for orbital periods in the
range $3-5$ days characteristic of hot Jupiters is only seen at high
metallicity. In addition, high eccentricities, possibly indicative of
dynamical interactions, are also predominantly seen at high
metallicity. This indicates multi-planet systems in which dynamical
interactions leading to close orbiters occur at high metallicity, and
that disc-driven migration is favoured at low metallicity.

Finally, misalignments between stellar equators and orbital planes may
not require strong dynamical interactions.  Several mechanisms may
produce misalignments between the protoplanetary disc and the
equatorial plane of its host star (see section~\ref{sec:type2ei}).
Another possibility is that internal processes within the star, such
as the propagation of gravity waves in hot stars, lead to different
directions of the angular momentum vector in the central and surface
regions \citep[e.g.,][]{2012ApJ...758L...6R}.

\begin{figure}[!t]
 \epsscale{0.95}
 \plotone{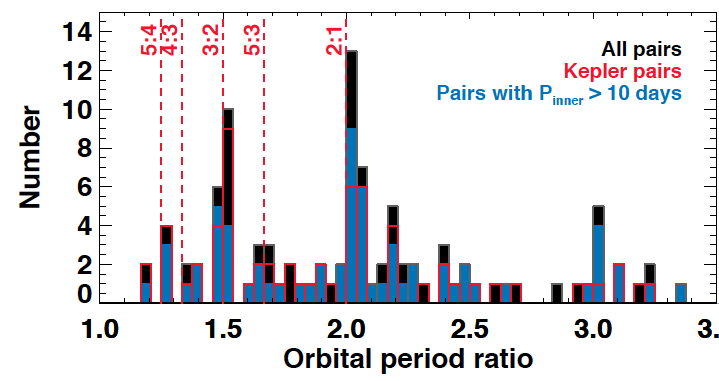}
\plotone {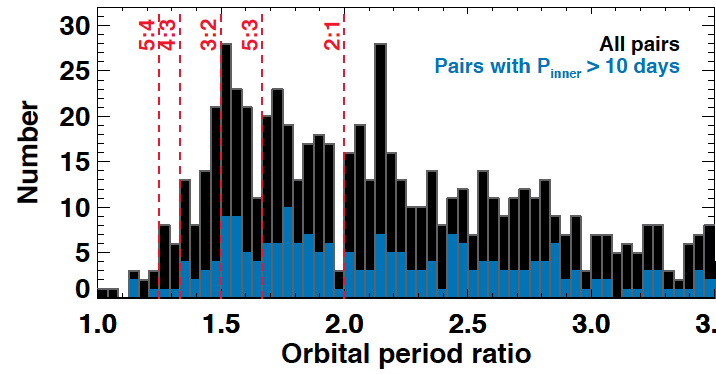}
 \caption{ \small  The period ratio histogram for confirmed exoplanets  
  is shown in the upper panel. The same histogram is shown in the 
  lower panel for Kepler candidates from Quarters 1 to 8. Vertical dashed 
  lines show the period ratio of a few mean-motion resonances. 
  Data were extracted from exoplanets.org.
  \label{fig:jp3}}
\end{figure}

\subsubsection{Low-mass planets}
\label{sec:johnlow}
\noindent 
The Kepler mission has discovered tightly packed planetary systems
orbiting close to their star.  Several have been determined to be
accurately coplanar, which is a signature of having formed in a
gaseous disc.  These include KOI 94 and KOI 25 \citep{Albrecht13}, KOI
30 \citep{Sanchis12}, and Kepler 50 and Kepler 55 \citep{Chaplin13}.
We remark that if one supposed that formation through in situ gas free
accumulation had taken place, for planets of a fixed type, the
formation time scale would be proportional to the product of the local
orbital period and the reciprocal of the surface density of the
material making up the planets \citep[e.g.,][]{PT06}. For a fixed mass
this is proportional to $r^{3.5}$. Scaling from the inner Solar
system, where this time scale is taken to be $\sim 3\times 10^8$ yr
\citep{1998Icar..136..304C}, it becomes $\lesssim 10^5$ yr for $r <
0.1$ AU. Note that this time scale is even shorter for more massive
and more compact systems. This points to a possible formation during
the disc lifetime, although the formation process should be slower and
quieter in a disc, as protoplanets are constrained to be in
non-overlapping near circular orbits. Under this circumstance,
disc-planet interactions cannot be ignored.
  
Notably, \citet{Lissauer11} found that a significant number of Kepler
multiplanet candidate systems contain pairs that are close to
first-order resonances.  They also found a few multi-resonant
configurations.  An example is the four planet system KOI-730 which
exhibits the mean motion ratios 8:6:4:3.  More recently,
\citet{Steffen13} confirmed the Kepler 60 system which has three
planets with the inner pair in a 5:4 commensurability and the outer
pair in a 4:3 commensurability.  However, most of the tightly packed
candidate systems are non-resonant.

The period ratio histogram for all pairs of confirmed exoplanet
systems is shown in the upper panel of Fig. \ref{fig:jp3}.  This shows
prominent spikes near the main first-order resonances.  However, this
trend is biased because many of the Kepler systems were validated
through observing transit timing variations, which are prominent for
systems near resonances.  The lower panel of this figure shows the
same histogram for Kepler candidate systems announced at the time of
writing (Quarters 1-8). In this case, although there is some
clustering in the neighbourhood of the 3:2 commensurability and an
absence of systems at exact 2:1 commensurability, with there being an
overall tendency for systems to have period ratios slightly larger
than exact resonant values, there are many non resonant systems.

At first sight, this appears to be inconsistent with results from the
simplest theories of disc-planet interactions, for which convergent
migration is predicted to form either resonant pairs \citep{Papa05} or
resonant chains \citep{CN06}.  However, there are features not
envisaged in that modelling which could modify these results, which we
briefly discuss below. These fall into two categories: (i) those
operating while the disc is still present, and (ii) those operating
after the disc has dispersed. An example of the latter type is the
operation of tidal interactions with the central star, which can cause
two planet systems to increase their period ratios away from resonant
values \citep{Papa11,2012ApJ...756L..11L,2013AJ....145....1B}.
However, this cannot be the only process operating as
Fig. \ref{fig:jp3} shows that the same period ratio structure is
obtained for periods both less than and greater than $10$ days. We
also note that compact systems with a large number of planets can be
close to dynamical instability through the operation of Arnold'
diffusion \citep[see, e.g., the analysis of Kepler 11
by][]{Miga12}. Thus it is possible that memory of the early history of
multi-planet systems is absent from their current configurations.

When the disc is present, stochastic forcing due to the presence of
turbulence could ultimately cause systems to diffuse out of resonance
\citep[e.g.,][]{PBH11,Rein12b}. When this operates, resonances may be
broken and period ratios may both increase and decrease away from
resonant values.  \citet{PRK13} employed stochastic fluctuations in
order to enable lower order resonances to be diffused through under
slow convergent migration. In this way, they could form a 7:6
commensurability in their modelling of the Kepler 36 system.

\begin{figure}[!t]
 \epsscale{0.85}
 \plotone{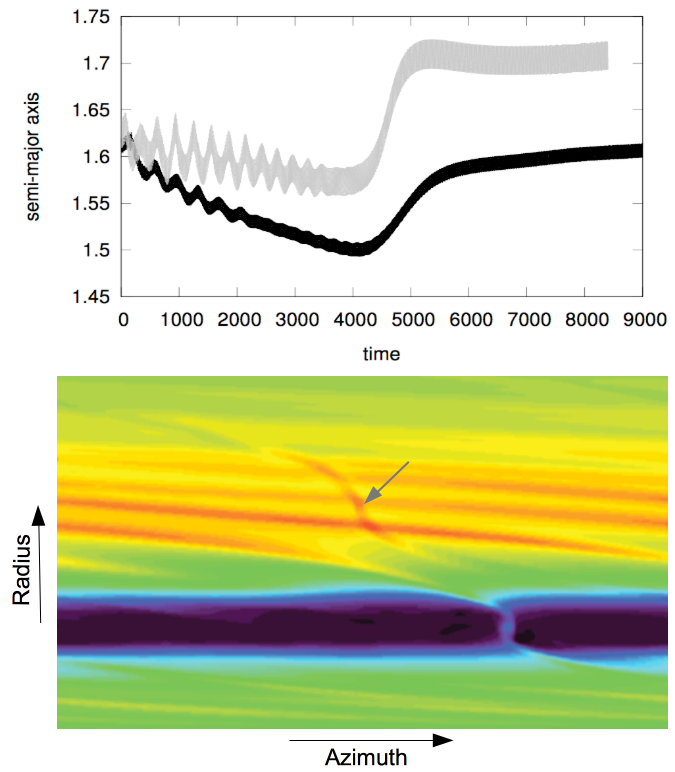}
 \caption{ \small Top: semi-major axis evolution of a $5.5M_{\oplus}$
   super-Earth orbiting exterior to an inwardly
     migrating giant planet of 1 $M_{\rm J}$ (dark line) or 2 $M_{\rm
     J}$ (light grey line) that is initially located
     at $a=1$. Time shown in x-axis is $2\pi$ times the
   initial orbital period of the
     giant planet. Bottom: disc's surface density for the one Jupiter
   mass case. The super-Earth's location is spotted by a grey
   arrow. Adapted from \citet{PPS12}.
   \label{fig:jp2}}
\end{figure}
Another mechanism that could potentially prevent the formation of
resonances in a disc involves the influence of each planet's wakes on
other planets. The dissipation induced by the wake of a planet in the
coorbital region of another planet causes the latter to be effectively
repelled. This repulsion might either prevent the formation of
resonances, or result in an increase in the period ratio from a
resonant value. \citet{PPS12} considered a $5.5M_{\oplus}$ super-Earth
migrating in a disc towards a giant planet. The upper panel in
Fig.~\ref{fig:jp2} shows the time evolution of the super-Earth's
semi-major axis when orbiting exterior to a planet of one Jupiter or
two Jupiter masses. We see that the super-Earth's semi-major axis
attains a minimum and ultimately increases. Thus a resonance is not
maintained.  The lower panel shows the disc's surface density at one
illustrative time: the super-Earth (grey arrow) feels a head wind from
the outer wake of the hot Jupiter, which leads to the super Earth
being progressively repelled. This mechanism could account for the
observed scarcity of super-Earths on near-resonant orbits exterior to
hot Jupiters.

\begin{figure}[!t]
  \epsscale{0.9} \plotone{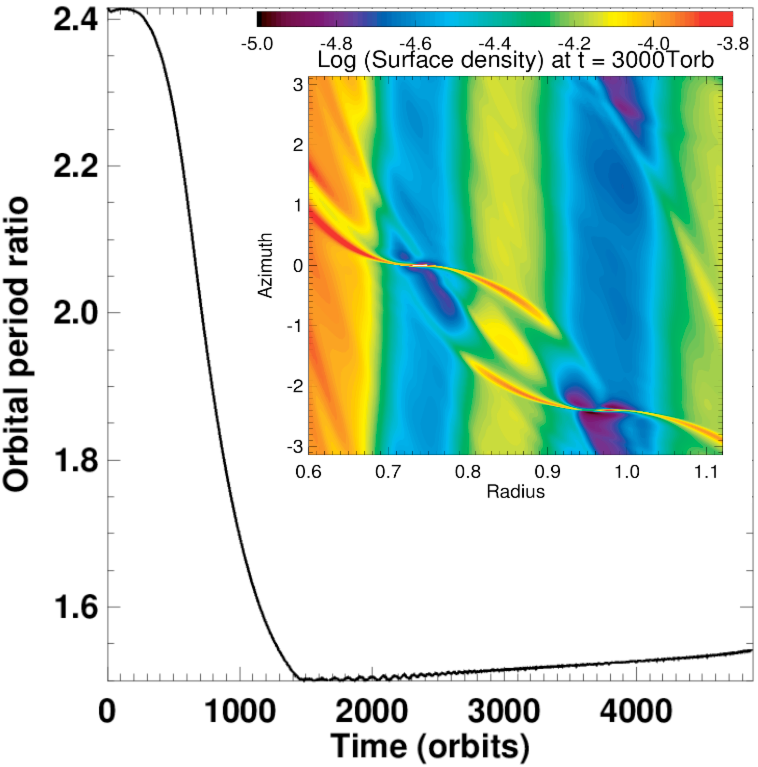}
  \caption{ \small Evolution of the orbital period ratio between an 
  $13 M_{\oplus}$ outer planet and an $15 M_{\oplus}$ inner  
  planet. The period ratio increases from the $3$:$2$ resonant ratio 
  due to wake-planet interactions. These interactions are visible in 
  the disc's surface density shown in the inset image. Adapted from 
  \citet{BP13}.
  \label{fig:jp4}}
\end{figure}

A similar effect was found in disc-planets simulations with two
partial gap-opening planets by \citet{BP13}. This is illustrated in
Fig.~\ref{fig:jp4}. The orbital period ratio between the planets
initially decreases until the planets get locked in the 3:2
mean-motion resonance. The period ratio then increases away from the
resonant ratio as a result of wake-planet interactions. The inset
panel shows a density contour plot where the interaction of the
planets with each other's wakes is clearly seen. Divergent evolution
of a planet pair through wake-planet interactions requires some
non-linearity, and so will not work with pure type I migration
($\mathcal{P} \gg 1$), but not so much that the gaps become totally
cleared \citep{BP13}. This mechanism works best for partial
gap-opening planets ($\mathcal{P} \sim$ a few, $q \sim h^3$), which
concern super-Earth to Neptune mass planets in discs with aspect ratio
$h \lesssim 3\%$ (expected in inner disc regions), or Saturn-mass
planets if $h \sim 5\%$. These results show circumstances where
convergent migration followed by attainment of stable strict
commensurability may not be an automatic consequence of disc-planet
interactions. Wake-planet interactions could explain why near-resonant
planet pairs amongst Kepler's multiple candidate systems tend to have
period ratios slightly greater than resonant.

\subsection{Planets on long-period orbits}
\label{sec:long}
\noindent In the past decade, spectacular progress in direct imaging
techniques have uncovered more than 20 giant planets with orbital
separations ranging from about ten to few hundred AU. The four planets
in the HR~8799 system \citep{Marois10} or $\beta-$Pictoris b
\citep{Lagrange10} are remarkable examples. The discoveries of these
{\it cold Jupiters} have challenged theories of planet formation and
evolution. We review below the mechanisms that have been proposed to
account for the cold Jupiters.

\subsubsection{Outward migration of planets formed further in?}
\noindent In the core-accretion scenario for planet formation, it is
difficult to form Jupiter-like planets in isolation beyond $\sim 10$
AU from a Sun-like star \citep[][and see the chapter by Helled et
al.]{Pollack96, IdaLin1}. Could forming Jupiters move out to large
orbital separations in their disc?  Outward type~I migration followed
by rapid gas accretion is possible, but the maximum orbital separation
attainable through type~I migration is uncertain (see
sections~\ref{sec:coro} and~\ref{sec:turb}).  Planets in the
Jupiter-mass range are expected to open an annular gap around their
orbit (see section~\ref{sec:gapopening}). If a deep gap is carved,
inward type II migration is expected. If a partial gap is opened,
outward type III migration could occur under some circumstances (see
section~\ref{sec:type3}), but numerical simulations have shown that it
is difficult to sustain this type of outward migration over long
timescales \citep{mp03,Peplinski3}.  It is therefore unlikely that a
{\it single} massive planet formed through the core-accretion scenario
could migrate to several tens or hundreds of AU.

It is possible, however, that a pair of close giant planets may
migrate outwards according to the mechanism described in
Section~\ref{sec:ms01}. For non-accreting planets, this mechanism
could deliver two near-resonant giant planets at orbital separations
comparable to those of the cold Jupiters \citep{Crida09}.  However,
this mechanism relies on the outer planet to be somewhat less massive
than the inner one. Joint outward migration may stall and eventually
reverse if the outer planet grows faster than the inner one
\citep{DAngeloMarzari}. Numerical simulations by these authors showed
that it is difficult to reach orbital separations typical of the cold
Jupiters.

\subsubsection{Planets scattered outwards?}
\noindent The fraction of confirmed planets known in multi-planetary
systems is about $1/3$ (see, e.g., exoplanets.org), from which nearly
$2/3$ have an estimated minimum mass (the remaining $1/3$ comprises
Kepler multiple planets confirmed by TTV, and for which an upper mass
estimate has been obtained based on dynamical stability; see for
example \cite{Steffen13}). More than half of the confirmed multiple
planets having a lower mass estimate are more massive than Saturn,
which indicates that the formation of several giant planets in a
protoplanetary disc should be quite common. Smooth convergent
migration of two giant planets in their parent disc should lead to
resonant capture followed by joint migration of the planet pair
\citep[e.g.,][]{Kley04}.  Dispersal of the gas disc may trigger the
onset of dynamical instability, with close encounters causing one of
the two planets to be scattered to large orbital separations
\citep{Chatterjee08}. A system of three giant planets is more prone to
dynamical instability, and disc-driven convergent migration of three
giant planets may induce planet scattering even in quite massive
protoplanetary discs \citep{MBS10,Moeckel12}. Planet scattering before
or after disc dispersal could thus be a relevant channel for
delivering one or several massive planets to orbital separations
comparable to the cold Jupiters'. It could also account for the
observed free-floating planets.

\subsubsection{Formation and evolution of planets by gravitational
  instability?}
\noindent Giant planets could also form after the fragmentation of
massive protoplanetary discs into clumps through the gravitational
instability (GI). The GI is triggered as the well-known Toomre-$Q$
parameter is $\sim 1$ and the disc's cooling timescale approaches the
dynamical timescale \citep{Gammie01, Rafikov05}. The later criterion
is prone to some uncertainty due to the stochastic nature of
fragmentation \citep{Paard12}. The GI could trigger planet formation
typically beyond 30 AU from a Sun-like star.

\begin{figure}[!t]
 \epsscale{0.9}
 \plotone{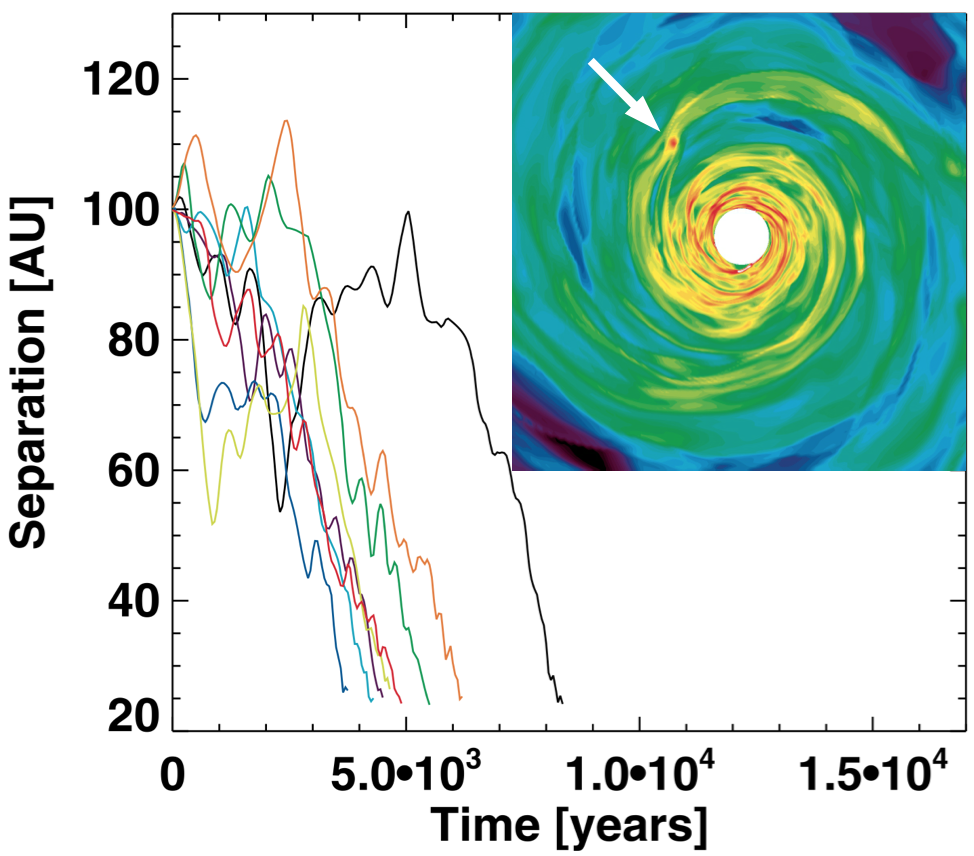}
 \caption{\small Semi-major axis evolution of a Jupiter-mass planet formed 
 in a self-gravitating turbulent disc at 100 AU from a Solar-mass star. Different 
 curves are for different simulations with varying initial conditions. The inset image 
 shows the disc's surface density in one of these simulations, and the 
 arrow spots the planet's location. Adapted from \cite{bmp11}.
  \label{fig:bmp}
}
\end{figure}
\par How do planets evolve once fragmentation is initiated? First,
GI-formed planets are unlikely to stay in place in their
gravito-turbulent disc. Since these massive planets form in about a
dynamical timescale, they rapidly migrate to the inner parts of their
disc, having initially no time to carve a gap around their orbit
\citep[][see Fig.~\ref{fig:bmp}]{bmp11, zhu12,Vorobyov13}.  These
inner regions should be too hot to be gravitationally unstable, and
other sources of turbulence, like the MRI, will set the background
disc profiles and the level of turbulence.  The rapid inward migration
of GI-formed planets could then slow down or even stall, possibly
accompanied by the formation of a gap around the planet's
orbit. Gap-opening may also occur if significant gas accretion occurs
during the initial stage of rapid migration \citep{zhu12}, which may
promote the survival of inwardly migrating clumps. Planet--planet
interactions, which may result in scattering events, mergers, or
resonant captures in a disc, should also play a prominent role in
shaping planetary systems formed by GI. The near resonant architecture
of the HR~8799 planet system could point to resonant captures after
convergent migration in a gravito-turbulent disc.

Furthermore, gas clumps progressively contract as they cool down.  As
clumps initially migrate inwards, they may experience some tidal
disruption, a process known as tidal downsizing. This process could
deliver a variety of planet masses in the inner parts of
protoplanetary discs \citep{Boley10,Nayak10}.

\subsection{Planet formation with migration: population synthesis and N-body
simulations}
\label{sect:migrat-obs}
\noindent
All of the disc-driven migration scenarios discussed in this review
have some dependence on the planet mass, so it is necessary to
consider the combined effects of mass growth and migration when
assessing the influence of migration on the formation of planetary
systems. Two approaches that have been used extensively for this
purpose are planetary population synthesis and N-body simulations,
both of which incorporate prescriptions for migration.

Population synthesis studies use Monte-Carlo techniques to construct
synthetic planetary populations, with the aim of determining which
combinations of model ingredients lead to statistically good fits to
the observational data \citep[orbital elements and masses in
particular, but in more recent developments planetary radii and
luminosities are also calculated as observables; see,
e.g.,][]{2012A&A...547A.112M}. In principle this allows the
mass-period and mass-radius relation for gaseous exoplanets to be
computed. Input variables that form the basis of the Monte-Carlo
approach include initial gas disc masses, gas-to-dust ratios, and disc
photoevaporation rates, constrained by observational data. The
advantages of population synthesis studies lie in their computational
speed and ability to include a broad range of physical processes. This
allows the models to treat elements of the physics (such as gas
envelope accretion, or ablation of planetesimals as they pass through
the planet envelope, for example) much more accurately than is
possible in N-body simulations. A single realisation of a Monte-Carlo
simulation consists of drawing a disc model from the predefined
distribution of possibilities and introducing a single low-mass
planetary embryo in the disc at a random location within a predefined
interval in radius. Accretion of planetesimals then proceeds, followed
by gas envelope settling as the core mass grows. Runaway gas accretion
to form a gas giant may occur if the core mass reaches the critical
value. Further implementation details are provided in the chapter on
planetary population synthesis by Benz et al. in this volume.

The main advantages of the N-body approach are that they automatically
include an accurate treatment of planet-planet interactions that is
normally missing from the `single-planet-in-a-disc' Monte-Carlo
models, they capture the competitive accretion that is inherently
present in the oligarchic picture of early planet formation, and they
incorporate giant impacts between embryos that are believed to provide
the crucial last step in terrestrial planet formation.  At present,
however, gas accretion has been ignored or treated in a crude manner
in N-body models. As such, population synthesis models can provide an
accurate description of the formation of a gas giant planet, whereas
N-body models are well-suited to examining the formation of systems of
lower mass terrestrial, super-Earth and core-dominated Neptune-like
bodies.

As indicated above, the basis of almost all published population
synthesis models has been the core-accretion scenario of planetary
formation, combined with simple prescriptions for type~I and type~II
migration and viscous disc evolution
\citep{2002MNRAS.334..248A,IdaLin1,2005A&A...434..343A,2009A&A...501.1139M}.
A notable exception is the recent population synthesis study based on
the disc fragmentation model \citep{2013MNRAS.432.3168F}.  Almost all
studies up to the present time have adopted type~I migration rates
similar to those arising from Eq.~(\ref{eqTL}), supplemented with an
arbitrary reduction factor that slows the migration. The influence of
the vortensity and entropy-related horseshoe drag discussed in
Sect.~2.1.2 has not yet been explored in detail, although a couple of
recent preliminary explorations that we describe below have appeared
in the literature.

\cite{2008ApJ...673..487I}, \cite{2009A&A...501.1139M} and
\cite{2009A&A...501.1161M} consider the effects of type~I and type~II
migration in their population synthesis models. Although differences
exist in the modelling procedures, these studies all conclude that
unattenuated type~I migration leads to planet populations that do not
match the observed distributions of planet mass and semimajor
axis. Models presented in \cite{2008ApJ...673..487I}, for example,
fail to produce giant planets at all if full-strength type~I migration
operates. Statistically acceptable giant planet populations are
reported for reductions in the efficiency of type~I migration by
factors of 0.01 to 0.03, with type~II migration being required to form
`hot Jupiters'.  With the type~II time scale of $\sim 10^5$ yr being
significantly shorter than disc life times, numerous giant planets
migrate into the central star in these models.  The survivors are
planets that form late as the disc is being dispersed (through viscous
evolution and photoevaporation), but just early enough to accrete
appreciable gaseous envelopes.  \cite{2009A&A...501.1139M} and
\cite{2009A&A...501.1161M} present models with full-strength type~I
migration that are able to form a sparse population of gas
giants. Cores that accrete very late in the disc life time are able to
grow to large masses as they migrate because they do not exhaust their
feeding zones. Type~I migration of the forming planetary cores in this
case, however, strongly biases the orbital radii of planetary cores to
small values, leading to too many short period massive gas giants that
are in contradiction of the exoplanet data.

The above studies focused primarily on forming gas giant planets, but
numerous super-Earth and Neptune-mass planets have been discovered by
both ground-based surveys and the Kepler mission
\citep[e.g.][]{2011arXiv1109.2497M,2011ApJ...736...19B,
  2013ApJ...766...81F}.  Based on 8 years of HARPS data, the former
publication in this list suggests that at least 50 \% of solar-type
stars hosts at least one planet with a period of 100 days or
less. Based on an analysis of the false-positive rate in Kepler
detections, \citet{2013ApJ...766...81F} suggest that 16.5 \% of FGK
stars have at least one planet between 0.8 and 1.25 R$_{\oplus}$ with
orbital periods up to 85 days.  These results appear consistent with
the larger numbers of super-Earth and Neptune-like planets discovered
by Kepler.  In a recent study, \cite{2010Sci...330..653H} performed a
direct comparison between the predictions of population synthesis
models with radial-velocity observations of extrasolar planets
orbiting within 0.25 AU around 166 nearby G-, K-, and M-type stars
(the $\eta_{\rm Earth}$ survey).  The data indicate a high density of
planets with $M_{\rm p}=$ 4 - 10 M$_{\oplus}$ with periods $<10$ days,
in clear accord with the discoveries made by Kepler. This population
is not present in the Monte-Carlo models because of rapid migration
and mass growth.  \cite{2010ApJ...719..810I} recently considered
specifically the formation of super-Earths using population synthesis,
incorporating for the first time a treatment of planet-planet
dynamical interactions.  In the absence of an inner disc cavity
(assumed to form by interaction with the stellar magnetic field) the
simulations failed to form systems of short period super-Earths
because of type~I migration into the central star. This requirement
for an inner disc cavity to halt inward migration, in order to explain
the existence of the observed short-period planet population, appears
to be a common feature in planetary formation models that include
migration. Given that planets are found to have a wide-range of
orbital radii, however, it seems unlikely that this migration stopping
mechanism can apply to all systems. Given the large numbers of planets
that migrate into the central star in the population synthesis models,
it would appear that such a stopping mechanism when applied to all
planet-forming discs would predict the existence of a significantly
larger population of short-period planets than is observed. This point
is illustrated by Fig.~7 which shows the mass-period relation for
planets with masses $0.1 \le M_{\rm p} \le 1$ M$_{\rm J}$ in the upper
panel and $10^{-3} \le M_{\rm p} \le 10^{-1}$ M$_{\rm J}$ in the lower
panel. Although a clustering of giant planets between orbital periods
3-5 days is observed, there is no evidence of such a pile-up for the
lower mass planets. This suggests that an inner cavity capable of
stopping the migration of planets of all masses may not be a prevalent
feature of planet forming discs.

N-body simulations with prescriptions for migration have been used to
examine the interplay between planet growth and migration. We
primarily concern ourselves here with simulations that include the
early phase of oligarchic growth when a swarm of Mars-mass embryos
embedded in a disc of planetesimals undergo competitive accretion. A
number of studies have considered dynamical interaction between much
more massive bodies in the presence of migration, but we will not
consider these here. Early work included examination of the early
phase of terrestrial planet formation in the presence of gas
\citep{2005AJ....130.2884M}, which showed that even unattenuated type
I migration was not inconsistent with terrestrial planet formation in
discs with a moderately enhanced solids abundance.  N-body simulations
that explore short period super-Earth formation and demonstrate the
importance of tidal interaction with the central star for disc models
containing inner cavities have been presented by
\cite{2007ApJ...654.1110T}.
\cite{2009MNRAS.392..537M,2010MNRAS.401.1691M} examined the formation
of hot super-Earth and Neptune mass planets using N-body simulations
combined with type~I migration (full strength and with various
attenuation factors).  The motivation here was to examine whether or
not the standard oligarchic growth picture of planet formation
combined with type I migration could produce systems such as Gliese
581 and HD 69830 that contain multiple short period super-Earth and
Neptune mass planets.  These hot and warm super-Earth and Neptune
systems probably contain up to 30 -- 40 Earth masses of rocky or icy
material orbiting within 1 AU.  The models incorporated a
purpose-built multiple time-step scheme to allow planet formation
scenarios in global disc models extending out to 15 AU to be explored.
The aim was to examine whether or not hot super-Earths and Neptunes
could be explained by a model of formation at large radius followed by
inward migration, or whether instead smaller building blocks of
terrestrial mass could migrate in and form a massive disc of embryos
that accretes {\it in situ} to form short period bodies. As such this
was a precursor study to the recent {\it in situ} models that neglect
migration of \citet{2013arXiv1301.7431H}.  The suite of some 50
simulations led to the formation of a few individual super-Earth and
Neptune mass planets, but failed to produce any systems with more than
12 Earth masses of solids interior to 1 AU.

\cite{2008Sci...321..814T} presented a suite of simulations of giant
planet formation using a hybrid code in which emerging embryos were
evolved using an N-body integrator combined with a 1D viscous disc
model. Although unattenuated type~I and type~II migration were
included, a number of models led to successful formation of systems of
surviving gas giant planets. These models considered an initial
population of planetary embryos undergoing oligarchic growth extending
out to 30 AU from the star, and indicate that the right combination of
planetary growth times, disc masses and life times can form surviving
giant planets through the core-accretion model, provided embryos can
form and grow at rather large orbital distances before migrating
inward.

The role of the combined vorticity- and entropy-related corotation
torque, and its ability to slow or reverse type~I migration of forming
planets, has not yet been explored in detail. The survival of
protoplanets with masses in the range $1 \le M_{\rm p} \le 10$
M$_{\oplus}$ in global 1D disc models has been studied by
\citet{2010ApJ...715L..68L}. These models demonstrate the existence of
locations in the disc where planets of a given mass experience zero
migration due to the cancellation of Lindblad and corotation torques
(zero-migration radii or planetary migration traps).  Planets have a
tendency to migrate toward these positions, where they then sit and
drift inward slowly as the gas disc disperses.  Preliminary results of
population synthesis calculations have been presented by
\cite{2011IAUS..276...72M}, and N-body simulations that examine the
oligarchic growth scenario under the influence of strong corotation
torques have been presented by \cite{2012MNRAS.419.2737H}. These
studies indicate that the convergent migration that arises as planets
move toward their zero-migration radii can allow a substantial
increase in the rate of planetary accretion. Under conditions where
the disc hosts a strongly decreasing temperature gradient,
\cite{2012MNRAS.419.2737H} computed models that led to outward
migration of planetary embryos to radii $\sim 50$ AU, followed by gas
accretion that formed gas giants at these large distances from the
star.  The temperature profiles required for this were substantially
steeper than those that arise from calculations of passively heated
discs, however, so it remains to be determined whether these
conditions can ever be realised in a protoplanetary disc.  Following
on from the study of corotation torques experienced by planets on
eccentric orbits by \citet{Bitsch10}, \cite{2012MNRAS.419.2737H}
incorporated a prescription for this effect and found that
planet-planet scattering causes eccentricity growth to values that
effectively quench the horseshoe drag, such that crowded planetary
systems during the formation epoch may continue to experience rapid
inward migration. Further work is clearly required to fully assess the
influence of the corotation torque on planet formation in the presence
of significant planet-planet interactions.

Looking to the future, it is clear that progress in making accurate
theoretical predictions that apply across the full range of observed
exoplanet masses will be best achieved by bringing together the best
elements of the population synthesis and N-body approaches. Some key
issues that require particular attention include the structure of the
disc close to the star, given its influence in shaping the
short-period planet population (see section~\ref{sec:short}).  This
will require developments in both observation and theory to constrain
the nature of the magnetospheric cavity and its influence on the
migration of planets of all masses. Significant improvements in
underlying global disc models are also required, given the sensitivity
of migration processes to the detailed disc physics.  Particular
issues at play are the roles of magnetic fields, the thermal evolution
and the nature of the turbulent flow in discs that sets the level of
the effective viscous stress.  These are all active areas of research
at the present time and promise to improve our understanding of planet
formation processes in the coming years.

\medskip
\section{\textbf{SUMMARY POINTS}}
\label{sec:summary}
\noindent The main points to take away from this chapter are
summarized below:
\begin{itemize}
\item Disc-planet interactions are a natural process that inevitably
  operates during the early evolution of planetary systems, when
  planets are still embedded in their protoplanetary disc. They modify
  all orbital elements of a planet.  While eccentricity and
  inclination are usually damped quickly, the semi-major axis may
  increase or decrease more or less rapidly depending on the
  planet-to-star mass ratio ($q$) and the disc's physical properties
  (including its aspect ratio $h$).
\item Planet migration comes in three main flavors. (i) Type I
  migration applies to low-mass planets ($\mathcal{P}\gg 1$, which is
  the case if $q \ll h^3$) that do not open a gap around their
  orbit. Its direction (inwards or outwards) and speed are very
  sensitive to the structure of the disc, its radiative and turbulent
  properties in a narrow region around the planet.  While major
  progress in understanding the physics of type~I migration has been
  made since PPV, robust predictions of its direction and pace will
  require more knowledge of protoplanetary discs in regions of planet
  formation. ALMA should bring precious constraints in that
  sense. (ii) Type II migration is for massive planets
  ($\mathcal{P}\lesssim 1$, or $q\geqslant q_{\rm min}$ given by
  Eq.~\ref{eq:qcrit}) that carve a deep gap around their
  orbit. Type~II migrating planets drift inwards in a time scale
  comparable to or longer than the disc's viscous time scale at their
  orbital separation. (iii) Type III migration concerns
  intermediate-mass planets ($q \sim h^3$) that open a partial gap
  around their orbit. This very rapid, preferentially inward migration
  regime operates in massive discs.
\item Planet-disc interaction is one major ingredient for shaping the
  architecture of planetary systems. The diversity of migration paths
  predicted for low-mass planets probably contributes to the diversity
  in the mass-semi-major axis diagram of observed
  exoplanets. Convergent migration of several planets in a disc could
  provide the conditions for exciting planets eccentricity and
  inclination.
\item The distribution of spin-orbit misalignments amongst hot
  Jupiters is very unlikely to have an explanation based on a single
  scenario for the large-scale inward migration required to bring them
  to their current orbital separations. Hot Jupiters on orbits aligned
  with their central star point preferentially to a smooth disc
  delivery, via type II migration, rather than to dynamical
  interactions with a planetary or a stellar companion, followed by
  star-planet tidal re-alignment.
\item Convergent migration of two planets in a disc does not
  necessarily result in the planets being in mean-motion resonance.
  Turbulence in the disc, the interaction between a planet and the
  wake of its companion, or late star-planet tidal interactions, could
  explain why many multi-planet candidate systems discovered by the
  Kepler mission are near- or non-resonant.  Wake-planet interactions
  could account for the observed scarcity of super-Earths on
  near-resonant orbits exterior to hot Jupiters.
\item Recent observations of circumstellar discs have reported the
  existence of cavities and of large-scale vortices in
  millimetre-sized grains.  These features do not necessarily track
  the presence of a giant planet in the disc. It should be kept in
  mind that the gaps carved by planets of around a Jupiter mass or
  less are narrow annuli, not cavities.
\item Improving theories of planet-disc interactions in models of
  planet population synthesis is essential to make progress in
  understanding the statistical properties of exoplanets. Current
  discrepancies between theory and observations point to uncertainties
  in planet migration models as much as to uncertainties in planet
  mass growth, the physical properties of protoplanetary discs, or to
  the expected significant impact of planet-planet interactions.
\end{itemize} 

\par\noindent 
{\it Acknowledgments}\, We thank Cornelis Dullemond
and the anonymous referee for their constructive reports.  CB was
supported by a Herchel Smith Postdoctoral Fellowship of the University
of Cambridge, SJP by a Royal Society University Research Fellowship,
JG by the Science and Technology Facilities Council, and BB by the
Helmholtz Alliance Planetary Evolution and Life.

\smallskip
\bibliographystyle{ppvi}

\end{document}